\documentclass[twocolumn,floatfix,prb,aps,amsmath, amssymb,superscriptaddress]{revtex4}
\usepackage{float}
\usepackage{graphicx}
\usepackage{amsmath}
\usepackage{amssymb}%This is for the symbol $\mathbb{Z}_2$.
\usepackage{color}

\newcommand{\ck}{\textcolor{blue}}

\begin{document}
\title{Pairing in Luttinger Liquids and Quantum Hall States}

\author{Charles L. Kane}
\affiliation{Department of Physics and Astronomy, University of Pennsylvania, Philadelphia, PA 19104}
\author{Ady Stern}
\affiliation{Department of Condensed Matter Physics, The Weizmann Institute of Science, Rehovot 76100, Israel}
\author{Bertrand I. Halperin}
\affiliation{Department of Physics, Harvard University, Cambridge, MA 02138}

\begin{abstract}

We study  spinless electrons in a single channel quantum wire
interacting through attractive interaction, and the quantum Hall states that may be constructed by an array of such wires. For a single wire the electrons may form two phases, the Luttinger liquid and the strongly paired phase. The Luttinger liquid is gapless to one- and two-electron excitations, while the strongly paired state is gapped to the former and gapless to the latter.
In contrast to the case in which the wire is proximity-coupled to an external superconductor, for an isolated wire there is no separate phase of a topological, weakly paired, superconductor. Rather, this phase is adiabatically connected to the Luttinger liquid phase. The properties of the  one dimensional topological superconductor emerge when the number of channels in the wire becomes large. The quantum Hall states that may be formed by an array of single-channel wires depend on the Landau level filling factors. For odd-denominator fillings $\nu=1/(2n+1)$, wires at the Luttinger phase form Laughlin states while wires in the strongly paired phase form bosonic fractional quantum Hall state of strongly-bound pairs at a filling of $1/(8n+4)$. The transition between the two is of the universality class of Ising transitions in three dimensions. For even-denominator fractions $\nu=1/2n$ the two single-wire phases translate into four quantum Hall states. Two of those states are bosonic fractional quantum Hall states of weakly- and strongly- bound pairs of electrons. The other two are non-Abelian quantum Hall states, which originate from coupling wires close to their critical point. One of these non-Abelian states is the Moore-Read state. The transition between all these states are of the universality class of Majorana transitions. We point out some of the properties that characterize the different phases and the phase transitions.

\end{abstract}

%\pacs{ }
\maketitle

\section{Introduction}

Luttinger liquid theory is a powerful tool for the theoretical study of interacting systems in one dimension, allowing for detailed analysis of their low energy properties\cite{haldane81jpc,haldane81prl}. Many attempts have been made to extend the theory for the study of systems of higher dimensions. A particularly successful
approach employs an array of quantum wires, each being described as a Luttinger liquid of spinless electrons, to construct two dimensional topological states of matter.  This coupled wire construction was originally formulated to describe Abelian fractional quantum Hall states\cite{kml2002}, and has been generalized to describe non-Abelian quantum Hall states\cite{teokane2014}, as well as other two and three dimensional topological phases\cite{neupert2014,sagi2014,santos2015,meng2015,sagi2015,huang2016,iadecola2016}.

While for systems of spinful electrons, or systems of more than one channel, Luttinger liquid theory is able to analyze instabilities towards
%phase transitions to
other phases, including superconductors or spin-gapped states, it does not find such instabilities for the minimal one dimensional system: a single channel quantum wire of spinless fermions.

In this work we use the Luttinger liquid as a starting point for exploring phases and phase transitions of a single-channel wire of spinless electrons that interact attractively. We have two goals. Within the  realm of one dimension, we are interested in the possible superconducting phases that are constructed in finite one dimensional systems, where the number of electrons is conserved. In particular, we are interested in the distinction between a Luttinger liquid, a topological superconductor and a non-topological superconductor in such systems.

Beyond the one dimensional realm, we are interested in using wires of attractively interacting electrons as building blocks for two dimensional fractional quantum Hall states. Here, we are motivated by the understanding that the most prominent series of non-Abelian quantum Hall states, the Read-Rezayi series\cite{mooreread1991,readrezayi1999}, are intimately related to clusters of electrons. In particular, the $\nu=5/2$ Moore-Read state is a paired state.  We seek a coupled wire construction of this state that incorporates the pairing physics and simplifies the construction of Ref. \onlinecite{teokane2014}, which involved an unnatural spatial modulation of the magnetic field.

Our study begins with a single channel quantum wire, continues with wires of many modes and then focuses on quantum Hall states formed by an array of single-mode wires. For single-mode wires, we analyze the strong pairing and weak pairing superconducting phases that occur for attractively interacting electrons. Interestingly, we find that the weakly paired phase is adiabatically connected to the Luttinger liquid phase formed by repulsively interacting electrons, but is separated by a phase transition from the strongly paired phase.
%The transition between the two phases is of the 2D Ising universality class, and
The two phases differ in their spectrum of single  electron excitations. While the strong pairing phase is gapped to single electrons and gapless to pairs of electrons, we find that the weak pairing phase, like the Luttinger liquid phase, is gapless to both. These results are different from those obtained in Mean Field Theory (MFT) of superconductivity, which is applicable when the superconductivity in the wire is induced by proximity coupling to an external bulk superconductor.

For multi-mode wires we discuss even-odd modulations of the ground state energy as a function of the number of electrons as well as single electron tunneling density of states, and contrast the results of the multi-mode wire to the case of a single-mode wire coupled by proximity to an external superconductor. We examine the way by which  the two systems become similar in the limit of a large number of modes.

For quantum Hall states formed of arrays of single mode wires, we focus on states in which the edge is composed of a single charged mode. We find two distinct phase diagrams as a function of the strength of the pairing. For odd-denominator filling factors $\nu=1/(2n+1)$ (with $n=0,1,2...$) there are two possible states - a strong pairing state, which is essentially a bosonic fractional quantum Hall state of filling $\frac{1}{4(2n+1)}$, and a Laughlin state. The two states differ in topological properties such as the quasi-particle charge and the ground state degeneracy on a torus. The transition between the two is of the 3D Ising universality class. For even-denominator filling factors $\nu=1/2n$ there are three types of states - strong pairing states, non-Abelian states of the Moore-Read type, and anistropic Abelian quantum Hall states, with defects that carry non-Abelian localized Majorana fermions.   In addition to the conventional Moore-Read state, our construction allows a related state that has counter-propagating charge and Majorana modes at the edge.  This state has the same topological order as the particle-hole symmetric Moore-Read state that has been recently discussed in the context of the theory of the particle-hole symmetric half filled Landau level\cite{son2015}.

We present  the physical picture that emerges from our study and a summary of our results in the next section, Sec. \ref{summary}. Following that, in Sec. \ref{sec:singlemode} we present our analysis of the single-mode wire. In Sec. \ref{qharrays} we discuss the wire construction of quantum Hall states. Finally, Sec. \ref{discussion} concludes the paper.

%The structure of the paper is as follows. In the next section (\ref{summary-picture}) we summarize our results and present the physical picture behind them. In Section....

\section{Physical picture and summary of results \label{summary}}

\subsection{Single wire}

We begin by considering a single-channel quantum wire with attractive interactions.    BCS Mean Field Theory (MFT) predicts two topologically distinct superconducting states, which both have a single particle energy gap.   In the paradigmantic Kitaev chain\cite{kitaev2001}, the transition between the trivial and the topological superconducting phases is driven by tuning the chemical potential across the band edge.
%The transition is in the 2D Ising universality class, and describes a pair of gapless counterpropagating Majorana modes.
In the topological phase there exist zero energy Majorana modes at the ends of the wire.   Domain walls between the trivial and topological phases also host Majorana modes, and the presence of four or more domain walls leads to a topological degeneracy in the ground state.

Recent work has examined the role of fluctuations in the superconducting order parameter\cite{sau2011,fidkowski2011} in number conserving one-dimensional superconductors\cite{KrausZoller,ChenBurnell}. An emphasis was given to the case of quantum wires with several channels.    It was found that in the weak pairing phase the finite-size splitting in the topological ground state degeneracy is suppressed only as a power law of the distance between Majorana modes, rather than the exponential behavior predicted by MFT. The power-law dependence is a consequence of back-scattering amplitudes that are introduced by impurities or  by non-uniformities (e.g., the nonuniformitty that gives rise to the existence of topological-trivial interfaces).  The exponent in the power-law depends on the interactions as well as the number of channels, and when the number of channels is large the exponent becomes large, effectively recovering the MFT behavior.      These works did not explicitly address the fate of the single particle energy gap, the low energy single particle tunneling density of states or the nature of the transition to other one dimensional phases.

In our analysis we go beyond MFT by describing the wire as a coexistence of two coupled fluids: charge $e$ fermions and charge $2e$ bosonic Cooper pairs.  The two are coupled by a non-quadratic pairing term that breaks a bosonic pair into two electrons and pairs two electrons into a bosonic pair. This term makes the Hamiltonian charge conserving, and couples the low energy fluctuations of the gapless mode, whose existence is guaranteed by translational invariance, to the superconducting pairs.  We will show that
this model has two phases, which we refer to as strong- and weak- pairing phases.    We find that in the strong pairing phase the single electron fluid is gapped as in the MFT.
In contrast, the weak paired phase is gapless to the introduction of single electrons.

The weak paired  phase can be viewed in two ways:   it can be considered to be a one dimensional topological superconductor with a fluctuating phase, or it can be considered to be ordinary single channel Luttinger liquid with attractive interactions.  These two pictures are in fact {\it equivalent}: the topological superconductor is adiabatically connected to the single channel Luttinger liquid, and ultimately to non-interacting electrons. Despite this equivalence, we will continue using the name ``weak pairing phase"  when pairing aspects are at the focus of our attention.

In MFT, there is a gap for adding single electrons to the bulk of a topological superconductor.  The same is true in our theory for ``bare" single electron operators  that do not couple to the bosonic sector.   However, we will show that when fluctuations in the charge $2e$ sector are accounted for there exist ``composite electron" operators, which carry a single electron charge and commute with the pairing term, leading to gapless charge $e$ excitations.
 The composite electron  operators involve scattering off of the finite wave vector density fluctuations in the charge $2e$ fluid, and transfer momentum to the bosonic fluid.   Alternatively, adding a composite electron is equivalent to adding a bare electron along with tunneling a superconducting vortex across the wire.   The vortex tunneling leads to a $2\pi$ phase slip in the superconducting order parameter.   In a topological superconductor, this process by itself leads to a change in the local fermion parity of the ground state resulting in an excited quasiparticle above the gap\cite{kitaev2001,fukane2009}.   Adding the bare fermion then couples the system back to the low energy sector with no quasiparticles excited.

We establish the equivalence between the weak paired superconductor and the Luttinger liquid by employing a unitary transformation that transforms the original charge-$e$ fermion mode and charge-$2e$ boson mode into a bosonic charge mode and a fermionic neutral mode.  In doing so, we generalize the conventional bosonization formulas to account for the possiblilty of a strong paired phase.  The bosonic charge mode resembles the bosonic mode of an ordinary single channel Luttinger liquid, while the neutral mode describes the degrees of freedom of a $1+1$ dimensional transverse field Ising model, where the ordered and disordered phases of the Ising model correspond to the weak and strong paired phases respectively.   In Section \ref{bnit} we will show that composite electron operators may be written in the form
\begin{equation}
\Psi_{R/L}^\dagger \sim (\sigma^x+i\sigma^y) e^{i(\varphi_\rho \pm \theta_\rho)}
\label{eq1}
\end{equation}
where $\varphi_\rho$ and $\theta_\rho$ are the bosonic fields in the charge sector, and $\sigma^{x,y}$ are spin variables characterizing the transverse field Ising neutral sector.   In the ordered phase of the Ising model (the weak paired phase), $\sigma^x$ has long range order and may be replaced by its expectation value $\langle\sigma_x\rangle$.   We therefore recover the conventional bosonization of a single channel Luttinger liquid.  In the disordered phase of the Ising model (the strong paired phase) $\sigma_{x,y}$ have short range correlations in space and time, which implies there is a gap for charge $e$ excitations.

As in the conventional theory of bosonization, there are additional composite charge $e$ fermion operators that involve $2k_F$ backscattering, leading to operators similar to (\ref{eq1}) with all odd integer multiples of $\theta_\rho$.   However, our generalized theory includes an additional class of charge $e$ operators with an {\it even} number of $\theta_\rho$'s.   For example, the ``bare" fermion  operator has the form,
\begin{equation}
\Psi_{0}^\dagger \sim (\gamma_1 + i \gamma_2) e^{i\varphi_\rho}
\end{equation}
While this operator has the ``wrong" number of $\theta_\rho$'s, it is accompanied by Majorana fermion operators $\gamma_{1,2}$ acting in the neutral sector.  These  are related to $\sigma^{x,y}$ by a Jordan-Wigner string.  These operators are usually unimportant at low energy because  $\gamma_{1,2}$ have a gap in both the ordered and disordered phases of the Ising model.  However $\gamma_{1,2}$ are gapless at the transition, as well as near a boundary that hosts a Majorana zero mode in the weak paired phase.

The picture we outline above also applies to a wire with multiple channels when an attractive interaction leads to a superconducting energy gap in all but a single collective charge mode.   For weak attractive interaction an odd number of channels form a weakly-paired superconductor, with the set of properties described above. In contrast,  an even number of channels form a strongly paired superconductor.   However, this even-odd dependence on the number of channels does not hold when the attractive interaction is not weak.   In fact, the weak- to strong- pairing transition that occurs with the opening of new channels when the interaction is weak is a particular case for which our analysis applies, but is not the general case.
In either phase, however, the wire is described by the generalized single channel Luttinger liquid theory described above, with a Luttinger parameter $K_\rho$ that depends on the number of channels.   When the number of channels is large $K_\rho$ is large.   In the limit $K_\rho\rightarrow \infty$ we recover the classical MFT behavior.

In MFT the transition between the weak and strong paired phases is a transition between two gapped states, and is in the 2D Ising universality class. At the transition a pair of counter-propagating Majorana modes described by $\gamma_{1,2}$ become gapless. For a charge conserving wire the transition is between two states that each have a gapless collective charge mode.  One may wonder whether the nature of the transition changes.      This problem has been studied in Refs. [\onlinecite{sitte,alberton}].  The coupling between the Ising variables and the gapless charge mode is found to be marginally irrelevant, but nonetheless changes the nature of the critical behavior at the transition because it leads to a strong renormalization of the velocities of the modes.  Depending on the coupling, there are two possible behaviors: either the transition is converted to a discontinuous first order transition, or the transition exhibits a continuous transition that resembles the Ising transition, but with a logarithmic renormalization of the velocity.

The distinction between the strong and weak paired phases can be probed experimentally in two ways: (1) the tunneling density of states at the bulk and at the end of the wire and (2) the dependence of the ground state energy of a finite system on the parity of the number of particles.     In the strong pairing phase both of these properties follow the MFT prediction.    There is a single particle energy gap in the tunneling density of states, and the ground state energy exhibits an even-odd effect, in which the energy difference for an odd and even number of electrons remains finite in the thermodynamic limit, as is observed in mesoscopic superconductors.

For the weak paired topological superconductor it is interesting to compare these properties with those of a single channel Luttinger liquid, as well as with non interacting electrons. Due to the composite electron operator, the tunneling density of states in the bulk of the weak paired phase vanishes like a power of energy.    This behavior is identical to an ordinary Luttinger liquid.    For a finite wire, the tunneling density of states at the end differs from that of the bulk.   As argued in Refs. [\onlinecite{sau2011,fidkowski2011}] the $\delta$-function peak at zero energy due to the Majorana mode predicted by MFT is replaced by a power law divergence in the tunneling density of states.   The same behavior arises in a single channel Luttinger liquid, where tunneling density of states at the end of the wire is different from that in the middle\cite{kanefisher1991}.   When the interactions are strong or the number of modes in the wire is large the exponent in the bulk becomes large, and we effectively recover the gap predicted by MFT.   In parallel, the exponent at the end approaches $-1$, and we recover the sharp peak characteristic of a Majorana mode.

In MFT the even-odd effect in the weak pairing phase depends on the boundary conditions.   For open boundary conditions the even-odd effect is exponentially small in the system size $L$ because the odd electron can be added to the Majorana end mode.   For periodic boundary conditions (a ring), however, there are no Majorana modes, and the even-odd effect is finite for large $L$.  Since fluctuations eliminate the single particle gap we expect this to be modified.   We predict that for open boundary conditions the even-odd effect remains  exponentially small in the system's size, while for periodic boundary conditions it vanishes only as $1/L$.    This distinction between periodic and open boundary conditions can be understood by considering the ground state energy of a single channel Luttinger liquid, and persists even for non interacting electrons. The difference between the two boundary conditions is most easily understood at the non-interacting level. For a wire with periodic boundary conditions the single particle spectrum is doubly degenerate, leading to an even-odd pattern in the ground state energy as a function of the number of electrons. This degeneracy is absent in a wire with open boundary conditions.

\subsection{Array of wires as a fractional quantum Hall state}

The coupled wire model constructs  quantum Hall states by connecting neighboring wires by tunneling in a way that mutually gaps a right-moving mode on one wire with a left-moving mode on its neighbor.   Provided the tunneling conserves momentum, this leads to a bulk gap, while chiral modes are left gapless on the edge.    The balance of momentum equates the momentum transferred to the tunneling charge by the Lorentz force with the total change in the electrons' momentum as a consequence of the tunneling.  For tunneling of charge $q$ across the inter-wire spacing $d$ in magnetic field $B$, the Lorentz force provides momentum $qBd$.   Fermionic Laughlin states are stabilized by a process in which this momentum is balanced by the momentum $2k_F(2n+1)$ associated with an electron tunneling from momentum state $k_F$ on one wire to momentum state $-k_F$ on the neighbor, while backscattering $n$ electrons from $+k_F$ to $-k_F$ on each wire (here $n=0,1,2...$).  The balance is satisfied at the Laughlin filling factors $\nu = 2k_F/qBd = 1/(2n+1)$.  When the constituents particles are bosons, the Laughlin states at $\nu=1/(2n+2)$ are formed when the tunneling boson applies a density operator of momentum $(n+1)2\pi n_b$ in the two wires involved.

In the present work we consider a wire construction in which the wires are in (i) the weak paired phase, (ii) the strong paired phase and (iii) the critical point between them.    For the weak paired phase we simply reproduce the construction described above for the Laughlin states\cite{kml2002}. This is consistent with the equivalence we draw of the weakly paired phase with the Luttinger liquid.  In the strong paired phase the single electrons are frozen, and the wires may be regarded as composed of bosons of charge $2e$.   Tunneling then leads to a sequence of strong paired quantum Hall states of charge $2e$ bosons at (electron) filling factors $\nu = 4/(2n)$ which have quasiparticles with charge $2e/2n$ with statistics angle $2\pi/(2n)$ and a degeneracy $2n$ on a torus.
Notable examples include $\nu=2$ with semionic charge-$e$  quasiparticles, $\nu=1$ with charge $e/2$ quasiparticles and $\nu=1/2$ with charge-$e/4$ quasi-particle\cite{halperin1983}, known in the literature as the $K=8$ state\ck{\cite{WenOverbosch}}.

For odd-denominator filling factors $\nu=1/(2n+1)$ there exist both ordinary Laughlin states (for weak paired wires) as well as strong paired states (for strong paired wires).   We will argue that the transition between them is in the 3D Ising universality class.

For even-denominator filling factors $\nu=1/(2n+2)$ ''composite electrons" cannot  tunnel in a way that conserves momentum. The wires are then coupled by pair tunneling, which gaps the gapless modes in the bulk. In the strong pairing regime a bosonic FQHE state is obtained. On the weak pairing side, gapless Majorana zero modes are left at the wires' ends. The ends may be coupled by single electron tunneling, in which case the localized Majorana end modes disperse into a pair of counter-propagating Majorana modes.  Defects in the array, e.g., wires that break or terminate at the bulk of the sample, carry localized zero energy Majorana modes which do not hybridize with the gapped bulk. These modes are static non-Abelian defects, of the type found at the ends of 1D topological superconductors.

When the individual wires are at the transition between weak- and strong- pairing, each wire carries both a gapless charged pair of modes and a gapless neutral pair of Majorana modes. It is then possible to construct an electron operator that allows for a single electron tunneling with momentum  conservation. This operator couples both to the charge and to the neutral modes. Together with pair tunneling that couples only to the charge mode, the bulk is gapped, and the edge remains to carry chiral charge and Majorana modes. The precise form of the single electron tunneling determines the relative direction of motion of the charge and Majorana modes on the edge. When they co-propagate, the resulting state is topologically identical to the Moore-Read state, which is a $p_x+ip_y$ superconductor of composite fermions. When the two edge modes counter-propagate, the resulting state is a $p_x-ip_y$ super-conductor of composite fermions. Its edge structure is symmetric to a particle-hole transformation. For electrons on a plane the $p_x-ip_y$ state is believed to be inferior in energy to the Moore-Read state, since the latter may be associated with a trial wave function (the ''Pfaffian" wave function) that is entirely within the lowest Landau level, while the former necessarily involves states from higher Landau levels. For weakly coupled wires, a limit which is far from that of electrons on a plane, it is possible to have situations in which the two states are comparable in energy. The starting point in which each wire is in a critical state allows also for a construction of the anti-Pfaffian, the particle-hole conjugate of the Moore-Read state. Since this state carries more than one charged edge mode, its construction requires inter-wire tunnel-coupling  that goes further than nearest-neighboring wires.

\section{Paired States in 1D \label{sec:singlemode}}

We begin with a simple model of spinless charge $e$ fermions with density $\rho_e \ll 1$ on a one dimensional lattice.  Consider the Hamiltonian
\begin{equation}
{\cal H} = \sum_i -t (c_i^\dagger c_{i+1} + c_{i+1}^\dagger c_i) + \sum_{p} V_p n_i n_{i+p}
\end{equation}
with first and second neighbor interactions $V_1$ and $V_2$, and $V_{p>2}=0$. For $V_1,V_2>0$, this system will be a Fermi gas with repulsive interactions, and at low energy will be a Luttinger liquid with gapless charge $e$ excitations.   For $V_1 < 0$,  however, the ground state will favor forming two particle bound states.   For $V_2> -2 V_1$ and small $t$ the system will form a gas of charge $2e$ bosons with repulsive interactions.   Thus, at low energy, the system will be a Luttinger liquid of charge $2e$ particles, and there will be a gap $|V_1|$ for creating charge $e$ particles.

We wish to develop a low energy theory that is capable of describing both of these phases, as well as the behavior of a transition between them.
Our approach is to consider a low energy ``two fluid model" consisting of a channel of charge $e$ fermions coexisting with charge $2e$ bosons.

\subsection{Two Fluid Model}

Consider a single wire described by a single fermionic channel with dispersion $E(k) = \epsilon_0+k^2/2m$ and average density $\bar\rho_f$ coupled to a Luttinger liquid of charge $2e$ bosons with average density $\bar\rho_b$.  The Hamiltonian density is
$
{\cal H} = {\cal H}^0_f + {\cal H}^0_b
$
where
\begin{equation}
{\cal H}^0_b = \frac{v}{2\pi}[K(\partial_x\varphi)^2 + \frac{1}{K}(\partial_x\theta)^2
]- 2\mu (\partial_x\theta/2\pi+\bar\rho_b)
\end{equation}
\begin{equation}
H^0_f =   \psi^\dagger (\epsilon_0 - \partial_x^2/2m - \mu) \psi + u (\psi \partial_x \psi e^{ 2 i \varphi}  + h.c.)
\label{h0f}
\end{equation}
Here $u$ is the pair tunneling between the fermion channel and the boson Luttinger liquid channel, which is necessarily p-wave.   $\partial_x\theta/2\pi$ describes the fluctuations in the boson density about $\bar\rho_b$, and $[\varphi(x),\theta(x')] = i \pi \Theta(x-x')$.
The chemical potential $\mu$ couples to the total charge density $\rho_e = \psi^\dagger\psi + 2\rho_b$.

 MFT treats $\varphi$ as a classical variable, in which case $H^0_f$ is a one dimensional version of the Read Green model of a $p$-wave superconductor.  It describes a transition between a trivial and topological one dimensional superconductor, as a function of $\epsilon_0$, where $\epsilon_0<0$ is the ``strong paired" phase, while $\epsilon_0>0$ is the ``weak paired", or topological phase, which exhibits Majorana modes at the end.  In mean field theory there is a gap for adding a charge $e$ particle in {\it both} the strong and weak paired phases.   In order to correctly describe the correlation functions for a charge $e$ particle in both phases it is necessary to go beyond the mean field theory.

The low energy fluctuations in the boson density have important contributions near wave vectors $q_n = 2\pi n \bar\rho_b$,
\begin{equation}
\rho_b(x) = \bar\rho_b + \sum_n \rho_{n}(x)
\end{equation}
The long wavelength boson density fluctuation is $\rho_0(x) = \partial_x\theta(x)/\pi$.  The density wave at $q \sim 2\pi \bar\rho n$ has a phase modulated by $\theta$,
\begin{equation}
\rho_n(x) \propto e^{i n (2\pi \bar\rho_b x + \theta(x))}
\end{equation}
Importantly, the operators $\rho_n$ are local operators.

The bare fermion operator $\psi$ is a local operator, but it is gapped when $\epsilon_0 \ne 0$ because of the pairing term.   However, the local composite electron operators
\begin{equation}
\Psi_{+,n}^\dagger(x) \propto \psi^\dagger \rho_n
\label{psi+}
\end{equation}
and
\begin{equation}
\Psi_{-,n}^\dagger(x) \propto \psi e^{2i\varphi} \rho_n
\label{psi-}
\end{equation}
are {\it not} necessarily gapped because $e^{\pm i n \theta}$ introduces a $2\pi n$ phase slip into the superconducting phase $2\varphi$.  It is well known that when the phase across a topological insulator Josephson junction is advanced by $2\pi$ there is a level crossing that results in an excited quasiparticle, along with a shift in the Fermion parity of the ground state.   It follows that when $n$ is odd and the fermions are in the topological superconducting state, the fermion operator can anihilate this extra quasiparticle, returning the system to the low energy sector with no gapped quasiparticles.    Therefore, in the weak paired phase the operators $\Psi_{\pm,\pm 1}$ create gapless excitations and have power law correlations characteristic of a charge $e$ Luttinger liquid.

\subsection{Bosonization near the Ising transition\label{bnit}}

In order to make contact with the more familiar bosonization of the charge $e$ Luttinger liquid, it is useful to introduce a transformation that decouples the fermions and bosons by effectively transferring the charge of the fermions to the bosons.  We perform a canonical transformation ${\cal H} \rightarrow U {\cal H} U^\dagger$ generated by
\begin{equation}
U = e^{i\int dx (\psi^\dagger\psi - \bar\rho_f)\varphi(x)}
\end{equation}
where $\bar\rho_f$ is the average fermion density.  Identifying the new boson and fermion fields as $\varphi_\rho$, $\theta_\rho$ and $\psi_\sigma$, this has the effect of transforming
\begin{eqnarray}
\psi(x) &\rightarrow& \psi_\sigma(x) e^{-i\varphi_\rho}, \\
\varphi(x) &\rightarrow& \varphi_\rho(x),  \\
\theta(x) &\rightarrow& \theta_\rho(x) + \pi(\bar\rho_f x + \int_x^{\infty} dx \psi_\sigma^\dagger \psi_\sigma).
\end{eqnarray}
Thus, $\partial_x\theta_\rho/\pi$ now describes the fluctuations in the total electron density about the average value $\bar\rho_e =\bar\rho_f + 2\bar\rho_b$.   The fermion field $\psi_\sigma$ is neutral.

The transformed Hamiltonian has three terms,
\begin{equation}
{\cal H} = {\cal H}_\rho + {\cal H}_\sigma + {\cal H}_{\rm int}.
\end{equation}
The term describing the charged degrees of freedom is
\begin{equation}
{\cal H}_\rho = \frac{v}{2\pi}[K_\rho(\partial_x\varphi_\rho)^2 + \frac{1}{K_\rho}(\partial_x\theta_\rho)^2
]- \mu (\partial_x\theta_\rho/\pi+\bar\rho_e),
\label{hrho}
\end{equation}
where we identify $K_\rho = K$.  The term describing the neutral degrees of freedom is
\begin{equation}
{\cal H}_\sigma = \psi_\sigma^\dagger(\epsilon_0 - \partial_x^2/2m)\psi_\sigma + i u (\psi_\sigma \partial_x \psi_\sigma +\psi_\sigma^\dagger\partial_x\psi_\sigma^\dagger)
\label{hsigmaf}
\end{equation}
and the interaction term is
\begin{equation}
{\cal H}_{\rm int} =  \frac{(\partial_x\varphi_\rho)^2}{2m}  \psi_\sigma^\dagger \psi_\sigma - \frac{i\partial_x\varphi_\rho}{m}  \psi_\sigma^\dagger\partial_x\psi_\sigma
-\frac{v \partial_x\theta_\rho}{K_\rho} (\psi_\sigma^\dagger \psi_\sigma-\bar\rho_f).  \label{hint}
\end{equation}

The second term, ${\cal H}_\sigma$,  can be recognized as the mean field Read Green model, which has a second order topological transition at $\epsilon_0 = 0$ that is in the 2D Ising universality class.   Since $\psi$ has dimension $1/2$ at the transition, the first two terms in ${\cal H}_{\rm int}$ have dimension $3$ and are strongly irrelevant.   The third term, however, with dimension $2$ is marginal, and affects the critical behavior\cite{sitte,alberton}.
%In Appendix \ref{RG} we present a renormalization group analysis that shows that the first two terms in ${\cal H}_{\rm int}$ are strongly irrelevant at this transition and may safely be ignored.
%The third term, however, has an important effect on the critical behavior.   In the following subsection we will argue that it converts the transition to a first order transition.
Before describing the critical behavior we will examine the correspondence between the phases and critical behavior of our model with ${\cal H}_{\rm int} = 0$ and the corresponding behavior of the Ising model.    Our motivation for doing this is that away from the transition the gap in $\psi_\sigma$ renders ${\cal H}_{\rm int}$ irrelevant.
%Moreover, we will argue that when $K_\rho$ is large (which occurs, for example, in a quantum wire with many channels) the corrections to the critical behavior are weak, so that the second order Ising critical point is an appropriate starting point.
Moreover, we will find the ${\cal H}_{\rm int}=0$ limit a useful starting point in Section \ref{qharrays} when we consider 2 dimensional gapped phases constructed from coupled wires, which have a gapped charge mode.

The connection between (\ref{hsigmaf}) and the Ising model is well known.   It is convenient to view the fermion fields $\psi_\sigma$ as the continuum limit of lattice fermion fields $\psi_j$, which may be written as the Jordan-Wigner transformation of lattice spin variables
$\sigma_j^\pm = \sigma_j^x \pm i \sigma_j^y$
via
\begin{equation}
\sigma_j^\pm = \psi_j^\pm e^{i \pi \sum_{i>j} \psi_i^+ \psi_i^-}.
\label{jordanwigner}
\end{equation}
Here we write $\psi^+ \equiv \psi_\sigma^\dagger$ and $\psi^- \equiv \psi_\sigma$.  The Hamiltonian (\ref{hsigmaf}) is then the continuum limit of a 1+1 dimensional transverse field Ising model,
\begin{equation}
{\cal H}_I = \sum_j  h \sigma_j^z - J \sigma_j^x \sigma_{j+1}^x
\label{hising}
\end{equation}
with $\epsilon_0 = 2(h-J)$.
The weak paired phases $\epsilon_0 < 0$ corresponds to the ordered phase of the Ising
model with a non zero order parameter $\langle \sigma^x \rangle \ne 0$.  The strong paired
phase $\epsilon_0 > 0$ is the disordered phase with $\langle \sigma^x \rangle = 0$.

The local composite electron creation operators given in (\ref{psi+},\ref{psi-}) can now be written as
\begin{equation}
\Psi_{\pm,n}^\dagger \propto \xi_{n}^\pm  e^{i\varphi_\rho} e^{ i n( k_F x + \theta_\rho)}
\label{newbosonize1}
\end{equation}
where
$k_F \equiv \pi\bar\rho_e$ comes from the total electron density and
\begin{equation}
\xi_{n}^\pm = \psi^\pm  e^{ i n\pi\int_x^\infty \psi^\dagger\psi}.
\end{equation}
From (\ref{jordanwigner}) it is clear that when $n$ is even the Jordan Wigner string has no effect, while
when $n$ is odd, the Jordan Wigner string converts the fermion variable into the Ising variable.
We thus have,
\begin{equation}
\xi^\pm_{n}  = \left\{\begin{array}{ll}
\sigma^\pm = (\sigma^x \pm i \sigma^y)/2 & n \ {\rm odd}, \\  \\
\psi^\pm  = (\gamma^1 \pm i \gamma^2)/\sqrt{2} & n \ {\rm even}.
\end{array}\right.   \label{newbosonize2}
\end{equation}
where $\sigma^{x,y}$ are the continuum limit of Pauli spin matrices $\sigma^{x,y}_j$ defined on each lattice site, while $\gamma^{1,2}$ are Majorana fermion operators that form the continuum limit of the lattice fermion operators $\psi_j^{\pm}$.

Eqs. (\ref{newbosonize1}) and (\ref{newbosonize2}) can be viewed as a generalization of the conventional bosonization formulas to the vicinity of the pairing transition.   Deep in the weak paired phase, the Ising spins have long range order, so that in the terms with odd $n$ they may be replaced by their non-zero expectation value $\langle \sigma^\pm \rangle = \langle \sigma^x\rangle $.  These terms give the conventional bosonization formula for a charge $e$ Luttinger liquid, in which the integer coefficients of $\varphi_\rho$ and $\theta_\rho$ have the same parity.  The usual left and right moving fermions $\Psi_R$ and $\Psi_L$ are given by the $n=\pm 1 $ operators.   Note, however, that in the strong paired phase $\langle \sigma^\pm\rangle = 0$, and the Ising spins have expenontially decaying correlations.   It follows that in the strong paired phase there are no charge-$e$ operators with long range correlations, reflecting an energy gap for charge-$e$ particles.   Charge $2e$ particles, created by $e^{2i \varphi_\rho}$, do not have a gap.

The even $n$ terms, which have the ``wrong" number of $\theta_\rho$'s, involve $\psi_\sigma$ which has a gap everywhere except in the vicinity of the Ising transition.   Near the transition, $\Psi_{\pm,0}$ couples to the gapless Majorana modes present at the transition.   Expressed in terms of the Majorana operators, Eq. \ref{hsigmaf} becomes
\begin{equation}
{\cal H}_\sigma = \gamma^T \left[ -i u \partial_x \tau^z + (\epsilon_0 - \partial_x^2/2m) \tau^y \right] \gamma
\end{equation}
where the Pauli matrices $\tau^a$ act on $\gamma = (\gamma_1, \gamma_2)^T$.    At the transition $\epsilon_0=0$, $\gamma_1$ and $\gamma_2$ can be recognized as right and left moving chiral Majorana modes.      Non zero $\epsilon_0$ introduces a mass term that couples $\gamma_1$ and $\gamma_2$ leading to an energy gap.

The spectral properties of single electrons can be summarized by the imaginary time local single particle Green's function ${\cal G}(\tau) = \langle T_\tau[\psi_e(x,\tau)\psi_e^\dagger(x,0)]\rangle$, where in general the electron operator $\psi_e^\dagger$ is a sum over all possible composite operators,
\begin{equation} \label{composite}
\psi^\dagger_e = \sum a_{+,n} \Psi_{+,n}^\dagger + a_{-,n} \Psi_{-,n}^\dagger,
\end{equation}
with coefficients $a_{\pm,n}$ that depend on the interactions beteen the charge $e$ and charge $2e$ fluids.   It follows that
\begin{equation}
{\cal G}(\tau) = \sum_n  \frac{1}{\tau^{\frac{1}{2}(K_\rho^{-1} + n^2 K_\rho)}} g_{n,\sigma}(\tau).
\end{equation}
Each term factorizes into a Luttinger liquid-like power law in the charge sector times a contribution $g_{n,\sigma}(\tau)$ from the neutral sector.   The latter depend crucially on whether $n$ is even or odd.   The behavior in the weak and strong paired phases can analyzed by considering the low energy behavior near the Ising transition $\epsilon_0 \sim 0$, where the neutral sector should exhibit scaling behavior.

For even $n$, $g_{n \ {\rm even},\sigma}$ describes the correlation function of 1D Majorana fermions with a mass gap $|\epsilon_0|$.   Since $\psi$ has dimension $1/2$ at the transition we expect
\begin{equation}
g_{n \ {\rm even},\sigma}(\tau,\epsilon_0)  \sim {1\over \tau} F_e(\epsilon_0 \tau).
\end{equation}
The scaling function satisfies $F_e(x\rightarrow 0) \sim 1$, reflecting the gapless Majorana mode at the transition.   In the opposite limit $F_e(x\rightarrow \pm \infty) \sim  e^{-|x|}$, reflecting the mass gap proportional to $|\epsilon_0|$ on either side of the transition.

For odd $n$, $g_{n \ {\rm odd},\sigma}$ describes the correlation function of the Ising spin operator $\sigma_x$.   Since $\sigma_x$ has dimension $1/8$ at the transition we expect
\begin{equation}
g_{n \ {\rm odd},\sigma}(\tau,\epsilon_0) \sim {1\over \tau^{1/4}} F_o(\epsilon_0\tau)
\end{equation}
The scaling function satisfies $F_o(x\rightarrow 0) \propto 1$, describing the behavior exactly at the transition.   The behavior for $x\rightarrow \pm \infty$ reflects the spin correlations on either side of the transition.   On the ordered side, for $\epsilon_0>0$, we expect $\langle \sigma_x(\tau)\sigma_x(0)\rangle \sim \epsilon_0^{2\beta}$ independent of $\tau$, with $\beta=1/8$.   Thus, $F_o(x \rightarrow \infty) \sim x^{1/4}$.   On the disordered side, $\epsilon_0<0$, the spin correlations decay exponentially in $\tau$ with a correlation length $\xi \propto |\epsilon_0|^{-\nu}$ with $\nu=1$.   Therefore, $F_o(x\rightarrow -\infty) \sim e^{- |x|}$.

The behavior described above can be summarized as follows:  In the strong paired phase ${\cal G}(\tau) \sim e^{-\epsilon_0\tau}$, reflecting a gap, while in the weak paired phase the electron operator is dominated by $n=\pm 1$, giving ${\cal G}(\tau)  \sim 1/\tau^{(K_\rho + K_\rho^{-1})/2}$.   At the Ising critical point the $n=0$ and $n=\pm 1$ terms compete, and we have ${\cal G}(\tau) \sim 1/\tau^{{\rm Min}(K_\rho^{-1}/2 + 1/4, K_\rho/2 + K_\rho^{-1}/2+ 1)}$.   Thus, for $K_\rho < 3/2$ the $n=1$ term dominates, while for $K_\rho > 3/2$ the $n=0$ term dominates.   As discussed in the following section, however, this critical behavior will be modified by the interaction with the charge mode.

Finally, we note that in the weak paired state the fermion sector has a topologically non trivial gap.   This means that there exist zero energy Majorana zero modes at the ends of the wire.  Therefore, at the ends $\Psi_{\pm,0}^\dagger \sim \psi^\pm e^{i\varphi_\rho}$ is not gapped, and exhibits long range temporal correlations with ${\cal G}(\tau) \sim \tau^{-K_\rho^{-1}}$.   This reflects the well known fact that the exponent for tunneling into the {\it end} of a Luttinger liquid differs from the exponent for tunneling into the middle of a Luttinger liquid\cite{kanefisher1991}.

\subsection{Critical Behavior \label{criticalbehavior}}

The effect of the marginal interaction $\lambda \partial_x\theta_\rho \psi_\sigma^\dagger \phi_\sigma$ (with $\lambda = v/K_\rho$) on the critical behavior of the Ising transition has been analyzed in Refs. [\onlinecite{sitte, alberton}].   These works performed a renormalization group analysis that showed how $\lambda$, as well as the velocities $u$ and $v$ are renormalized at low energy at the transition.
We write here the RG equations in a scheme in which space and time are renormalized at identical rates and the coefficient of the $(\partial_x\varphi_\rho)^2$ term in Eq. (\ref{hrho}), namely the product $vK_\rho$, is held fixed.  The RG equations are then,
\begin{eqnarray}
\frac{d\log{v}}{dl}=-\frac{K_\rho \lambda^2}{4uv} \\
\frac{d\log{K_\rho}}{dl}=\frac{K_\rho \lambda^2}{4uv}\\
\frac{d\log{u}}{dl}=-\frac{K_\rho \lambda^2}{(u+v)^2}\\
\frac{d\log{\lambda}}{dl}=-\frac{K_\rho \lambda^2(2u+v)}{2u(u+v)^2}
\label{rgeqns}
\end{eqnarray}
These equations were found to result in two distinct types of flows. For the first type, $u, v$ and $\lambda$ flow to zero with $u/v=1$.  This fixed point describes a continuous transition in which the fermions and bosons are decoupled but their velocities both vanish logarithmically at low energy. In the second type of flow the boson velocity $v$ renormalizes to zero faster than $u, \lambda$, and flows to zero at a finite $\ell$.  This signals a first order transition that resembles a transition between the degenerate ground states associated with the Peierls instability.  The separatrix between the two types of low is not known precisely.

When the initial conditions are such that $\lambda\ll v$ and $u\ll v$ the transition is expected to be continuous, with a flow of the first type. While we do not provide a detailed estimate of the microscopic parameters, we will make two observations. First, in the situation where superconductivity is induced into a semi-conducting wire by proximity to metallic one-dimensional channels (see, e.g., Albrecht et al. \cite{Albrecht2016}), we expect $v\gg u$ since $v$ originates from the metal while $u$ originates from the superconductivity induced in the semi-conductor.  Second, in our analysis above the initial value of $\lambda$ is given by $v/K_\rho$. When the number of channels $N_{ch}$ is large we expect $K_\rho\propto N_{ch}$, such that the initial value of $\lambda\rightarrow 0$. In this case, Eqs. (\ref{rgeqns}) imply that the effect of renormalization becomes significant only at exponentially low energy. However, this analysis is modified if inter-channel Coulomb interaction is included. In that case, we find that $K_\rho, v$ are both proportional to $N_{ch}^{1/2}$ in the limit $N_{ch}\rightarrow\infty$. The initial values of the parameters are then not necessarily small, which complicates the determination of the RG flow.

\subsection{Distinguishing characteristics of the two phases}

We now discuss two physical characteristics that distinguish the weak-paired/Luttinger liquid phase from the strongly paired phase in a one dimensional wire. We also explore the case in which the wire has a large number of channels.

\subsubsection{Even-odd modulation of the ground state energy as a function of the electron number N}

We define the energy alternation $\Delta E(L)$, for a system of length $L$, by
\begin{equation}
\Delta E(L) = 2 \sum_{N=0}^\infty e^{i\pi n} E_N w(N)
\end{equation}
where $E_N$ is the ground state energy of a system  with $N$ fermions in the length $L$, and $w(N)$ is a smooth weight function, with a maximum at a value $N=n_0 L$ and a width which is large compared to unity but small compared to $n_0 L$. For example, $w(N)$ might be a Gaussian with a second moment equal to $n_0 L$.  We shall be interested in the behavior of $\Delta E$ as $L$ becomes large, with the mean fermion density $n_0$ held fixed.

$\Delta E(L)$ samples the even-odd modulation of the ground state energy over the range of $w(N)$, and is rather insensitive to the smooth variation of $E_N$. In MFT, for a wire with Majorana end modes it may be interpreted as the energy splitting due to interactions between the Majorana modes at the two ends of the wire, when such modes exist. It should not be confused with the second difference
\begin{equation}
\Delta^2 (E_N) \equiv E_{N+1} - 2E_N + E_{N-1} ,
\end{equation}
with $N \approx n_0 L$.
For a  particle conserving system in the topological superconducting state, with short range interactions, the   value of $\Delta^2 (E_N)$  will only go to zero as $1/L$, for large $L$, regardless of the number of channels, as the second difference samples the smooth dependence of $E_N$ on $N$.
For all phases there is a smooth contribution to the ground state energy of the form $E_c(N-N_0)^2/2$ which accounts for charging energy. This contribution determines $\Delta^2(E_N)=E_c\propto \frac{1}{L}$ but does not lead to an even-odd effect and gives a negligible contribution to $\Delta E(L)$.

Focusing on $\Delta E(L)$ we note that for a strong-paired superconductor there is another contribution $(-1)^N\Delta$, originating from the gap to having an unpaired electron. This leads to $\Delta E(L)$ that is independent of $L$.
The weak paired state, however, does not have a gap for single electrons, but rather behaves as a Luttinger liquid described by Eq. (\ref{hrho}).   The dependence of the ground state energy on parity depends on the boundary conditions.   For open boundary conditions, there is just a single chiral channel that reflects back and forth from the ends.   For a segment of length $L$ the energy is then,
\begin{equation}
E_N^{segment} = \frac{2\pi v}{L K_\rho} (N-N_0)^2.
\label{ensegment}
\end{equation}
This gives a negligible contribution to $\Delta E(L)$, reminiscent of the absence of an even-odd effect in a topological superconductor due to the Majorana end modes.    In contrast, for a ring geometry, there are independent left and right moving chiral modes with integer charges $N_{R/L}$, with $N=N_R+N_L$.   The energy for a ring of circumference $L$ is then,
\begin{equation}
E_N^{ring} = \frac{2\pi v}{L}\left(\frac{1}{K_\rho} (N_R + N_L - N_0)^2 + K_\rho (N_R-N_L)^2\right).
\label{enring}
\end{equation}
The ground state therefore has an even-odd modulation $\Delta E(L) = 2\pi v K/L$.

We note that this behavior of the even-odd effect persists even for non interacting electrons.   In this case in the segment geometry the single particle states are non degenerate and evenly spaced, leading to (\ref{ensegment}), while in the ring geometry every state (except at $k=0$) is doubly degenerate, leading to (\ref{enring}).

It is interesting to compare these conclusions to the conclusions of MFT, in which a quantum wire is proximity coupled
to an ideal infinite bulk superconductor.  In that case, the electron number $N$ is not a good quantum number. Then, the ground state of the wire, at a given chemical potential will be a superposition of states with different values of $N$, but a fixed number parity.  When the wire is in a topological state and the geometry is that of a segment the number parity of the ground state will alternate between even and odd as $L$ is increased, but the energy difference between even and odd parity should decrease exponentially with $L$. This would not happen for a ring, in which $\Delta E(L)$ would stay independent of $L$, since the MFT weakly paired phase is gapped to single electrons.

%\ck{i moved this here: still needs to be rewritten...  For a clean system, this splitting is determined by the MFT-looking Eq. (\ref{hsigmaf}) to be exponentially small in $L$. The splitting will be modified by disorder or non-uniformity in the wire. Those allow for back-scattering terms which couple $e^{i\theta_\rho}$ to the neutral fermion $\psi_\sigma$, and change the splitting to decay algebraically with $L$.}

\subsubsection{Tunneling density of states}

A quantity of particular interest is the spectral density $A(\varepsilon ,x)$ for tunneling an electron into the wire at a position $x$ and energy $\varepsilon$, measured relative to the Fermi energy.  Within MFT, if the position $x$ is close to the end of a semi-infinite wire  in the weak-pairing topological state,  there will be a peak in spectral density at very low energies, which is associated with a localized Majorana mode at the wire end.  In the case of the hybrid wire, coupled to an infinite superconductor, this peak should be a delta-function at zero energy. The amplitude of the delta function will fall off exponentially as $x$ moves away from the end of the wire, and there will be an energy gap about zero energy, where the spectral density is zero for any position $x$.

 For a wire with conserved particle number, for $x$ near and end of the wire, $A(x, \varepsilon)$  will have only power-law divergence at $\varepsilon=0$.  The finite spectral density at $\varepsilon \neq 0$ occurs because the injected electron will necessarily produce phonon-like excitations of the charge density in the wire, which exist down to arbitrarily low energy in a semi-infinite wire.  Although the amplitude of the zero-energy singularity will  decrease with increasing $x$, there should not be a hard gap in the spectral density, even for $x$ in the middle of an infinite wire. In the weak pairing phase there will always be some weight at low energies, in the middle of the wire,  though this weight will typically  decrease rapidly with the number of channels in the wire.

The tunneling density of states is closely related to the single particle Green's function discussed in Section \ref{bnit}.   We find the spectral density for tunneling an electron into a point near the end of a semi-infinite wire should have the form, for $\varepsilon \to 0$,
 \begin{equation}
 \label{Aend}
 A(x, \varepsilon) \sim C(x) | \varepsilon| ^{ \beta} ,
 \end{equation}
 \begin{equation}
 \beta = K_\rho^{-1} - 1 ,
 \end{equation}
where $K_{\rho}$ is the Luttinger parameter defined in Eq. (\ref{hrho}) above.
For a multichannel wire, the prefactor $C$ should fall off exponentially with distance from the end for short distances, and fall off as a power law for large distances.  By contrast, the spectral density for tunneling at a point in the bulk of the wire,  infinitely far from  an end,
should have the form
\begin{equation}
\label{Abulk}
A_{\rm{bulk}}(\varepsilon) \sim D |\varepsilon| ^\alpha ,
\end{equation}
where
\begin{equation}
\alpha = \frac {K_\rho + K_\rho^{-1}}{2} - 1,
\end{equation}
and $D$ is a constant of proportionality.
For a wire with attractive interactions, we expect to find $K_\rho > 1$, so both $\alpha$  and $\beta$  are positive, so that the spectral density at the end of the wire  should diverge for $\varepsilon \to 0$, while it vanishes in this limit in the bulk.  The expressions for these power laws have precisely the same form as in the well-known case of single channel spinless Luttinger liquid, where, typically, $K_\rho$ is not very different from unity.  In contrast, for a  the multichannel wire with weak attractive interactions, we expect to find
\begin{equation}
K_\rho  \sim \kappa N_{\rm {ch}}
\end{equation}
where $N_{\rm{ch}}$ is the number of channels and $\kappa$  is a number slightly larger than unity.
Thus the exponent $\alpha$ will become large,  when  $N_{\rm{ch}}$ is large, and $\beta$ will be close to -1, so that the right-hand side of (\ref{Aend}) will be close to a delta-function at zero energy.

We remark that there will be a constant  of proportionality on the right-hand- side of Eq (\ref{Abulk}), which should, itself, become small in the case of a large number of channels with weak interactions.  This is because the bare fermion creation  operator $\psi^{\dagger}  (x)$  does not couple directly to low energy excitations. Instead we must employ an operator of the form $\Psi^\dagger \sim \psi^{\dagger} \eta$, similar to the operators in Eqs. (\ref{psi+},\ref{psi-}), where $\eta$ is an operator that produces a phase slip of strength $\pm 2 \pi$ in the superconducting phase $\varphi$.  For wire with $N_{\rm{ch}} = 2n+1$, the operator $\eta$ will involve excitation of $n$ particles and $n$ holes in $2n$ different channels. Thus  the operator $\Psi^{\dagger}$ will appear only at order $n$ in perturbation theory, and should therefore be small,  proportional to the $n$-th power of the ratio between the  interaction strength and the Fermi energy, for weak interactions.  (In Ref [\onlinecite{sau2011}] these higher-order terms were overlooked, and it was incorrectly suggested that  there should be a hard gap in the spectral weight for a wire with three or more channels.)

In contrast to the weak paired phase, the strong paired phase has a gap for  single particle excitations.  The coefficient $D$ in (\ref{Abulk}) is zero, and $A(\varepsilon) = 0$ for $\varepsilon < \Delta$.
%Since the transition between weak and strong paired phases is first order, we expect a discontinuous transition, where $D$ ($\Delta$) drops discontinuously to zero in the strong (weak) paired phases.
However, since two particle tunneling is allowed in the strong paired phase there will remain a nonzero tunneling conductance $G$, due to a process analogous to Andreev reflection, which is suppressed by both a higher power of energy as well as a higher power in the bare electron tunneling matrix element.   While in the weak paired phase $G \sim t^2 A(\varepsilon) \sim t^2 \varepsilon^{K_\rho/2+K_\rho^{-1}/2 -1}$ (where $\varepsilon$ is the larger of temperature or voltage), we find that in the strong paired phase
$G \sim t^4 \varepsilon^{2(K_\rho + K_\rho^{-1})+2}$.

Both the weak- and strong-pairing states are one-dimensional  versions of a superconductor, in the sense that they have  superconducting order parameters with quasi-long-range order.   Specifically, the correlation functions for  operators that inject a pair of electrons at a point $x$ and remove a pair at point $x^\prime$ will fall off as a power of $| x - x^\prime |$, with an exponent that becomes small as the number of channels becomes large.

In a multichannel wire with weak interactions between the electrons, there is a difference between an even and an odd number of channels.  For an odd number of channels, with weak attractive interactions, the ground state is predicted to be a one-dimensional version of a topological superconductor, with  low-energy properties that coincide with the weak-pairing state discussed above\cite{sau2011,fidkowski2011}.  For a wire with an even number of channels and weak attractive interactions, the ground state is expected to be effectively a one-dimensional version of a non-topological superconductor, with a finite energy difference between states of  even and odd number parity, and an energy gap for adding a single electron to the wire. The low-energy properties of this state coincide with those of the strong-pairing situation discussed above.

\section{Paired Quantum Hall States in the Coupled Wire Model\label{qharrays}}

We now apply the formalism developed in Sec. (\ref{sec:singlemode}) to describe paired quantum Hall states.   Strong paired quantum Hall states can occur when electrons are strongly bound into pairs, which themselves form  fractional quantum Hall states of bosonic charge $2e$ particles.   The simple bosonic Laughlin states occur at electron filling factors $\nu  = 4 \nu_{2e} = 4/m$, where $m$ is an even integer and $\nu = \rho_e h/(eB) = 4 \rho_{2e} h/(2eB)$.    For $\nu = 1/2$, the strong paired state is equivalent to charge $2e$ bosons at filling $\nu_{2e} = 1/8$.    Strong paired phases can also occur at other filling factors, such as $\nu = 1$ ($\nu_{2e} = 1/4$).

Here we will show how these states can be formulated within the coupled wire model.  We will begin by reviewing the coupled wire model, and then show how it is modified by incorporating pairing into the Luttinger liquids.   We will then demonstrate two applications of this technique.  First we will show that the transition between the conventional $\nu=1$ state and the strong paired $\nu=1$ state is in the 3D Ising universality class.  Second we will provide an explicit construction of the Moore-Read quantum Hall state, as well as related weak paired states at $\nu = 1/2$, and a new,  intrinsically anisotropic quantum Hall state at $\nu = 1/2$.

\subsection{Coupled Wire Model}

The coupled wire model provides an explicit formulation of fractional quantum Hall states that takes advantage of the power of Abelian bosonization for describing strongly interacting quantum systems.   The power of studying the anisotropic limit of quantum Hall states has long been known in the theory of the integer quantum Hall effect\cite{sondhi2001}.
A coupled wire construction for Abelian fractional quantum Hall states was introduced in Ref. [\onlinecite{kml2002}], and later generalized in several directions\cite{teokane2014,neupert2014,sagi2014,santos2015,meng2015,sagi2015,huang2016,iadecola2016}.

We will begin by considering an array of coupled wires that are each single channel  Luttinger liquids described by conventional bosonization.  This is equivalent to considering (\ref{newbosonize1}) and (\ref{newbosonize2}) for $n$ odd, and setting $\sigma^\pm$ to a constant.   In this case
there is only a single fermion operator, so we will omit the $\pm$ subscript on $\Psi_j$ in (\ref{newbosonize1}).  It is convenient to choose a periodic boundary condition of circunference $L$ for the 2D system in which wire $j$ at $x=L$ connects to wire $j+1$ at $x=0$.  Thus, the 2D system consists of a single wire that is ``wrapped on a spool" with $\Psi_j(x) \equiv \Psi(x+jL)$.   In this case, the anticommutation between fermion operators is automatically built in, so that additional Klein factors are not necessary.

Suppose there is a magnetic flux per unit length $b = \bar\rho_e\phi_0/\nu$ between the wires (here $\rho_e \equiv 2k_F/(2\pi)$ is the total electron density per wire, and $\phi_0 = h/e$ is the flux quantum).   Tunneling an electron between neighboring wires is associated with a phase $2\pi b x/\phi_0 = 2k_F x/\nu$.    The allowed momentum conserving tunneling terms are determined by comparing the magnetic field phase to the phase due to the momentum in the electron operators.

Consider electron tunneling operators of the form
\begin{equation}
\Psi_{j,m}^\dagger \Psi_{j+1,-m} e^{2\pi ib x/\phi_0} + h.c.
\end{equation}
for odd integer $m$, where  $j$ and $j+1$ enumerate the wires.   When the filling factor is $\nu = 1/m$ the oscillating factor $e^{2\pi i b x/\phi_0}$ is exactly cancelled by the phase $e^{2i m k_F x}$ in the $\Psi$'s.     Expressed in bosonized form, the tunneling term then has the form,
\begin{equation}
{\cal H} = {\cal H}_\rho + \sum_j  t_1 \cos \Theta_{\rho,j+1/2} .
\end{equation}
where
\begin{equation}
\Theta_{j+1/2} \equiv \varphi_{\rho,j}-\varphi_{\rho,j+1} + m(\theta_{\rho,j} + \theta_{\rho,j+1})
\end{equation}
Here ${\cal H}_\rho$ is a sum of terms of the form ({\ref{hrho}) as well as forward scattering interactions that couple the wires.   In the spirit of the coupled wire model, we assume that the forward scattering interactions are such that $t_1$ is relevant under the renormalization group and flows to strong coupling.   It then follows that the set of mutually commuting variables $\Theta_{\rho,j+1/2}$, which are defined on each link between wires, are pinned at an integer multiple of $2\pi$, resulting in a gapped quantum Hall state.   A kink in which $\Theta_{\rho, j+1/2}$ jumps by $2\pi$ corresponds to a charge $e/m$ Laughlin quasiparticle.

To describe a strongly paired quantum Hall state, it is tempting to consider  the momentum conserving tunneling of {\it pairs} of electrons between the wires, described by
\begin{equation}
{\cal H} = {\cal H}_\rho + t_2 \sum_j \cos 2\Theta_{\rho,j+1/2} .
\label{t2}
\end{equation}
This term pins $\Theta_{j+1/2}$ at a multiple of $\pi$.  A kink in which $\Theta_{\rho, j}$ jumps by $\pi$ has a charge $e/(2m)$ that is expected for a Laughlin quasiparticle of charge $2e$ bosons at filling $1/(4m)$.

However, having $t_2$ flow to strong coupling, with $t_1=0$ does {\it not} describe the strong paired phase because it is unstable to the perturbation $t_1$ which is still a local operator.   When $t_1 \ne 0$, the $\pi$ kink no longer connects degenerate ground states, so that the charge $e/(2m)$ quasiparticles are confined.    This phase is just the conventional $\nu=1/m$ Laughlin state.

To describe the paired quantum Hall states we need to augment the single- and pair- tunneling terms with intra-wire pairing. We will employ our generalized bosonization approach (\ref{newbosonize1}) and (\ref{newbosonize2}), which includes the $\sigma$ sector, and incorporates strong pairing into the 1D wires.   We thus consider
\begin{equation}
{\cal H} = {\cal H}_\rho + {\cal H}_\sigma + \sum_j V_{1j} + V_{2j}
\end{equation}
where ${\cal H}_\rho$ and ${\cal H}_\sigma$ are the single wire Hamiltonians (\ref{hrho}) and (\ref{hsigmaf}), and the tunneling terms have the form
\begin{equation}
V_{2j} =  t_2 \cos 2\Theta_{\rho,j+\frac{1}{2}},
\label{t2equation}
\end{equation}
\begin{equation}
V_{1j} = \left\{\begin{array}{ll}   \sum_{a,b}  t_{1,ab}  \sigma^{ a}_{j} \sigma^{ b}_{j+1} \cos \Theta_{\rho,j+\frac{1}{2}}.  & m \ {\rm odd} \\  \\
\sum_{a,b}  t_{1,ab}  \gamma^{ a}_{j} \gamma^{ b}_{j+1} \cos \Theta_{\rho,j+\frac{1}{2}}
.  & m \ {\rm even} \\
\end{array}\right.
\label{oneetunneling}
\end{equation}
Here we have chosen to express the four allowed tunneling terms involving $\Psi_{\pm,j,m}$ and $\Psi_{\pm,j+1,m}$ in (\ref{newbosonize1}) in terms of the Hermitian operators $\sigma^{x,y}$ or $\gamma^{x,y}$.    We again assume that pair tunneling between the wires, described by (\ref{t2}), is relevant, and leads to the pinning of $\Theta_{\rho, j+1/2}$ at multiples of $\pi$. The choice between the two lines of (\ref{oneetunneling}) is dictated by the need to cancel the magnetic field phase $e^{2\pi i b x/\phi_0}$.

 When $m$ is odd, in addition to the conventional $\nu=1/m$ Laughlin state, we will describe the strong paired Laughlin state as well as the critical behavior of the transition between the weak and strong paired phases.    When $m$ is even, strong pairing on the wires leads to a strong paired $\nu=1/m$ Laughlin state, while weak pairing on the wires leads to a novel anisotripic quantum Hall state.   When the wires are near the transition between the weak and strong paired phases, we find that coupling the wires leads to states we can identify with the Moore-Read state as well as generalizations of it.  For both even and odd $m$, the $V_{2j}$ terms gap the charge modes at the bulk, and hence suppress the effect of the intra-wire interaction Hamiltonian ${\cal H}_{\rm int}$ in Eq. (\ref{hint}). Thus, the modifications to the critical behavior that we found for a single wire in Subsection (\ref{criticalbehavior}) do not occur in arrays of coupled wires. As we see below, the phase transitions in the two dimensional arrays of wires are continuous second order transitions, and involve a closure of the gap to neutral excitations.}

We will consider the cases of odd and even $m$ separately in the following two sections.

\subsection{Pairing Transition for $\nu = 1$ (or $m$ odd)}

In this section we study the strong paired $\nu=1/m$ state for $m$ odd, as well as the critical behavior of the transition to the conventional $\nu=1/m$ Laughlin state.   The simplest case is $\nu = 1$.     To this end, we consider the strong coupling limit $t_2 \rightarrow \infty$.  In the ground state sector in which no quasiparticles are present we may set $\Theta_{\rho,j+1/2} = 0$.     We will first introduce a mapping to a 2+1 dimensional transverse field Ising model, which allows us to show that the transition is in the 3D Ising universality class.   We will then explore the scaling behavior near the transition, which can be probed by transport and thermodynamic measurements.

\subsubsection{3D Ising Transition}

When $m$ is odd, the Hamiltonian in the neutral sector involves (\ref{hsigmaf}) on each wire, as well as the tunneling terms due to $t_{1,ab}$ (the top line of Eq. (\ref{oneetunneling})).   Since the tunneling term is best described by the Ising variables, it is convenient to employ the Ising lattice-regularized description also for the neutral sector of the wires' Hamiltonians, Eq. (\ref{hising}) and reviewed in Appendix A.   This then leads to an anisotropic 2+1D lattice model of the form,
\begin{equation}
H = \sum_{i,j} h \sigma^z_{i,j} + J \sigma^x_{i,j} \sigma^x_{i,j+1}
+ t_{1,ab} \sigma^a_{i,j}\sigma^b_{i+1,j}
\end{equation}
where $\sigma^a_{i,j}$ describes the spin on the j'th site of the i'th chain.
If we consider the anisotropic limit $t_{1,ab} \rightarrow 0$ and bosonize the weakly coupled chains using the ``spool" boundary conditions discussed above, then we arrive at a model with precisely the same Jordan Wigner strings as the fermion model.

For simplicity consider first the case where $t_{1, xx}$ is the only nonzero term.  Then we have precisely an anisotropic 2+1D transverse field Ising model.   This model has two phases that are separated by a critical point in the 3D Ising model universality class.

The ``high temperature phase" of the Ising model, $\langle \sigma^a \rangle = 0$ is the strong paired phase.   In this phase there is a single particle energy gap - even at the edge.    However, the charge sector has a gapless chiral Luttinger liquid edge mode, which allows the low energy tunneling of pairs of electrons.
In this phase, there is no energy cost to having a $\pi$ kink in $\Theta_{\rho,j+1/2}$ because the single particle tunneling term is suppressed by the disordered Ising spins.   Thus, the charge $e/(2m)$ quasiparticle is deconfined, as expected in the strong paired phase.      If we have a pair of quasiparticles, then for the Ising sector, they are connected by a string of flipped bonds.    In the dual description of the Ising model, which has the form of an Ising gauge theory, the quasiparticles are bound to $\pi$ fluxes.   In the disordered phase of the gauge theory (equivalent to the ordered phase of the Ising model) the $\pi$ fluxes are deconfined.

In the low temperature phase of the Ising model, $\langle \sigma^x \rangle \ne 0$.   There remains an energy gap in the bulk, but now the single particle propagator is gapless at the edge.   This is the ordinary $\nu=1/m$ Laughlin state.   In this case, the $\pi$ kinks with charge $e/(2m)$ are confined because the string corresponds to a domain wall that costs an energy proportional to its length.

The other tunneling terms $t_{1, xy}$ and $t_{1,yy}$ will not modify the phases or critical behavior - at least when they are weak.   At the 3D Ising critical point there is a single relevant operator, and all remaining operators are irrelevant. Therefore, adding these terms as perturbations can only shift the location of the transition, but not modify it.

\subsubsection{Critical Behavior}

At the transition between the strong paired and conventional $\nu=1/m$ state, the charge gap remains finite, so this transition will not exhibit scaling in the longitudinal DC conductance that occurs in conventional quantum Hall transitions.   To access the critical behavior it is necessary to probe the neutral degrees of freedom that acquire long ranged correlations at the critical point.   In this section we discuss two quantities:   the heat capacity and the edge tunneling conductance.   We will argue that both quantities exhibit scaling behavior that is sensitive to the bulk 3D Ising transition.

The scaling of the heat capacity $C(T,\delta)$ with temperature $T$ and distance to the critical point $\delta$ can be deduced from dimensional analysis.   With an appropriate linear rescaling of space and time the critical point exhibits a Lorentz invariance.  The only length scale is the correlation length, which diverges as $\xi \sim \delta^{-\nu}$, where $\nu = .630$  is the correlation length exponent of the 3D Ising model.    The only energy scale is $\Delta \sim v/\xi \sim \delta^{\nu}$.  Since the heat capacity per unit area has units $L^{-2}$, precisely at the transition we must have $C(T,\delta=0)\propto T^2$, which is characteristic of any 2D Lorentz invariant system.   More generally it is expected to exhibit scaling behavior,
\begin{equation}
C(T,\delta) = T^2 f(\delta/T^{1/\nu})
\end{equation}
Since for fixed $\delta \ne 0$ there is an energy gap $\Delta$, we can deduce the asymptotic behavior
\begin{equation}
f(X\rightarrow \pm \infty) \propto e^{-|X|^\nu}.
\end{equation}
At a fixed low temperature we therefore expect $C(T,\delta)$ to exhibit a peak as a function of $\delta$ near $\delta=0$.    The peak is predicted to sharpen as the temperature is lowered and data from different temperatures should collapse to a single curve in a scaling plot.

Tunneling into the edge is another probe of the pairing transition.   On the weakly paired side the edge tunneling is described by the conventional Luttinger liquid theory, which for filling $1/m$ predicts a low temperature tunneling conductance due to tunneling of single electrons that scales as $G_1(T) \sim  t^2 T^{m-1}$, where $t$ is the electron tunneling matrix element.   On the strongly paired side there is a gap for tunneling single electrons.  The dominant contribution comes from tunneling pairs of electrons.    This leads to a tunnel conductance $G_2(T) \sim t^4 T^{4m + 2}$.   In the critical region the single electron tunneling $G_1(T,\delta)$ will exhibit critical behavior, while the two particle contribution will behave smoothly.   Here we will focus on $G_1(T,\delta)$ which dominates in the large barrier limit $t\rightarrow 0$.

The single electron tunneling density of states follows from the single particle Greens function, ${\cal G}(\tau) = \langle T_\tau[\Psi(\tau)\Psi^\dagger(0)]\rangle$, where $\Psi^\dagger \propto \sigma^+ e^{i\phi_\rho}$.     Here $e^{i\phi_\rho}$ adds a charge $e$ in the charge sector, while in the neutral sector it involves the Ising spins.    The charge sector operators exhibit the usual Luttinger liquid behavior,
\begin{equation}
\langle e^{i\phi_\rho(\tau)} e^{-i\phi_\rho(0)}\rangle \sim 1/\tau^{m}
\end{equation}
when the filling is $1/m$.
On the ordered side of the transition we expect $\langle \sigma_x \rangle \propto \delta^\beta$, while on the disordered side $\langle \sigma_x\rangle = 0$.   In the critical region, the correlation function exhibits scaling,
\begin{equation}
\langle \sigma_x(\tau) \sigma_x(0) \rangle =    {1\over \tau^{1+\eta}} g({\delta\tau^{1/\nu}})
\label{correlation}
\end{equation}
 where $g(X\rightarrow 0)\sim X^{2\beta}$ and $g(X\rightarrow \infty) \sim X^\nu$ and the exponents $\beta = .326$ and $\eta = .036$ are related by $2\beta = \nu(1+\eta)$.    This, in turn leads to scaling behavior in the tunneling conductance.    We find that near the transition the single particle tunneling conductance has the form
\begin{equation}
G_1(T,\delta) = T^{m + \eta} h({\delta\over{T^{1/\nu}}})
\end{equation}
where $h(X\rightarrow \infty) = X^{2\beta} = X^{\nu(1+\eta)}$, and $h(X\rightarrow -\infty) \sim e^{-c X^\nu}$.

\begin{figure}
\includegraphics[width=3.5in]{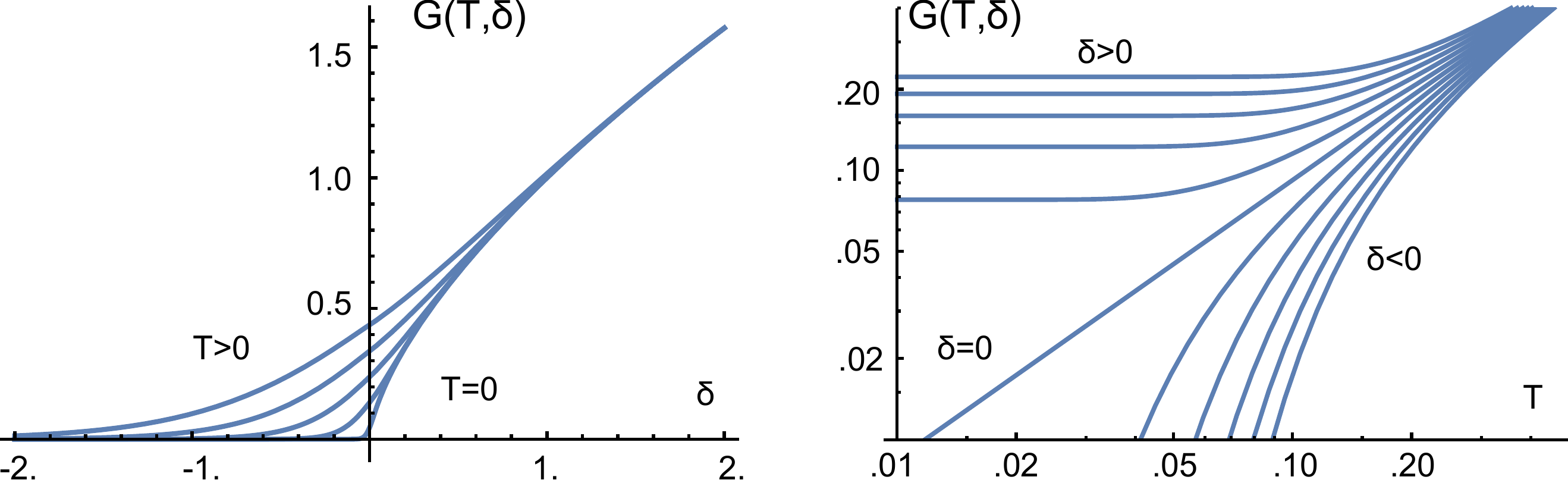}
\caption{Critical behavior of the edge tunneling conductance $G$ as a function of temperature $T$ and tuning parameter $\delta$ near the 3D Ising critical point.  (a) shows $G$ as function of $\delta$ for several values of $T$, highlighting the zero temperature behavior in which $G(0,\delta) \sim \theta(\delta)\delta^{2\beta}$.   (b) shows $G$ as a function of $T$ for several values of $\delta$, highlighting the behavior at the critical point $G(T,0) \sim T^{m+\eta}$.}
\label{scalingfig}
\end{figure}

In Fig. \ref{scalingfig} we show the predicted behavior for $m=1$ using a simple approximate scaling function $h(X)$ that interpolates between the known limits at $X=\pm \infty$.    Fig.\ref{scalingfig}a shows $G$ as a function of $\delta$ for several values of $T$.    At low $T$, it shows a sharp transition and grows as $\delta^{2\beta}$ for $\delta>0$.   At higher temperature the transition is rounded.    Fig.\ref{scalingfig}b shows a log-log plot of $G$ as a function of $T$ for several values of $\delta$.   For $\delta<0$ the low temperature conductance approaches a constant (or more generally $T^{m-1}$), while for $\delta>0$ it goes to zero exponentially.   Precisely at the transition $\delta=0$, $G(T) \propto T^{m+\eta}$.

In the above analysis we assumed that the critical behavior for correlation function (\ref{correlation}) at the Ising transition is the same on the boundary of the system as it is in the bulk.    The boundary exponents can in general be different.   This will modify the exponents and scaling functions, but the general form of the scaling behavior will remain.

\subsection{Paired Phases for $\nu=1/2$ (or $m$ even)}

We now consider the case of even  $m$, for which the simplest example is $\nu = 1/2$.    In this case there is no Laughlin state for fermions, but we expect to describe a strong paired state, as well as a non-Abelian Moore-Read state.
Our starting point is to have each wire at the pairing transition, so there exist gapless charge and Majorana modes on each wire.   We will then consider perturbations that couple the wires and lead to paired quantum Hall states.

When $m$ is even the single electron tunneling term is the second line of Eq. (\ref{oneetunneling}), and we can work with the 1D continuum fermion theory description of the wires' neutral sectors.   We again consider the strong coupling limit $t_2\rightarrow \infty$, and in the absence of quasiparticles set $\Theta_{\rho,j+1/2}=0$.
The Hamiltonian for the neutral sector then becomes ${\cal H} = {\cal H}_0 +{\cal H}_t$
\begin{equation}
{\cal H}_0 = \sum_{i} \gamma^{T}_i [ -i u \tau^z \partial_x + (\epsilon_0-\partial_x^2/2m)\tau^y] \gamma_i
\label{h000i}
\end{equation}
and
\begin{equation}
{\cal H}_t =  i \sum_i \gamma_i^{aT} t_{1,ab} \gamma_{i+1}^b .
\end{equation}
Fourier transforming we  may write
\begin{equation}
H = \gamma^T(-k) {\bf h}(k) \gamma(k)
\label{h2by2}
\end{equation}
with the $2\times 2$ matrix
\begin{eqnarray}
{\bf h}(k) = T_1 \sin k_y I + (u k_x + \Delta_1 \sin k_y)\tau^z + \nonumber \\
(\epsilon_0 + k_x^2/2m + T_2 \cos k_y)\tau^y + \Delta_2 \sin k_y \tau^x
\label{h000f}
\end{eqnarray}
where $T_1 = t_{1,11}+t_{22}$, $T_2 = t_{12}-t_{21}$, $\Delta_1 = t_{11} - t_{22}$ and
$\Delta_2 = t_{12}+t_{21}$.   $T = T_1+iT_2$ can be interpreted as the complex tunneling amplitude $T\psi_j^\dagger \psi_{j+1}$, while $\Delta = \Delta_1 + i\Delta_2$ can be interpreted as the complex pariring term with $p_y$ symmetry, $\Delta \psi_j \psi_{j+1}$.

\begin{figure}
\includegraphics[width=2in]{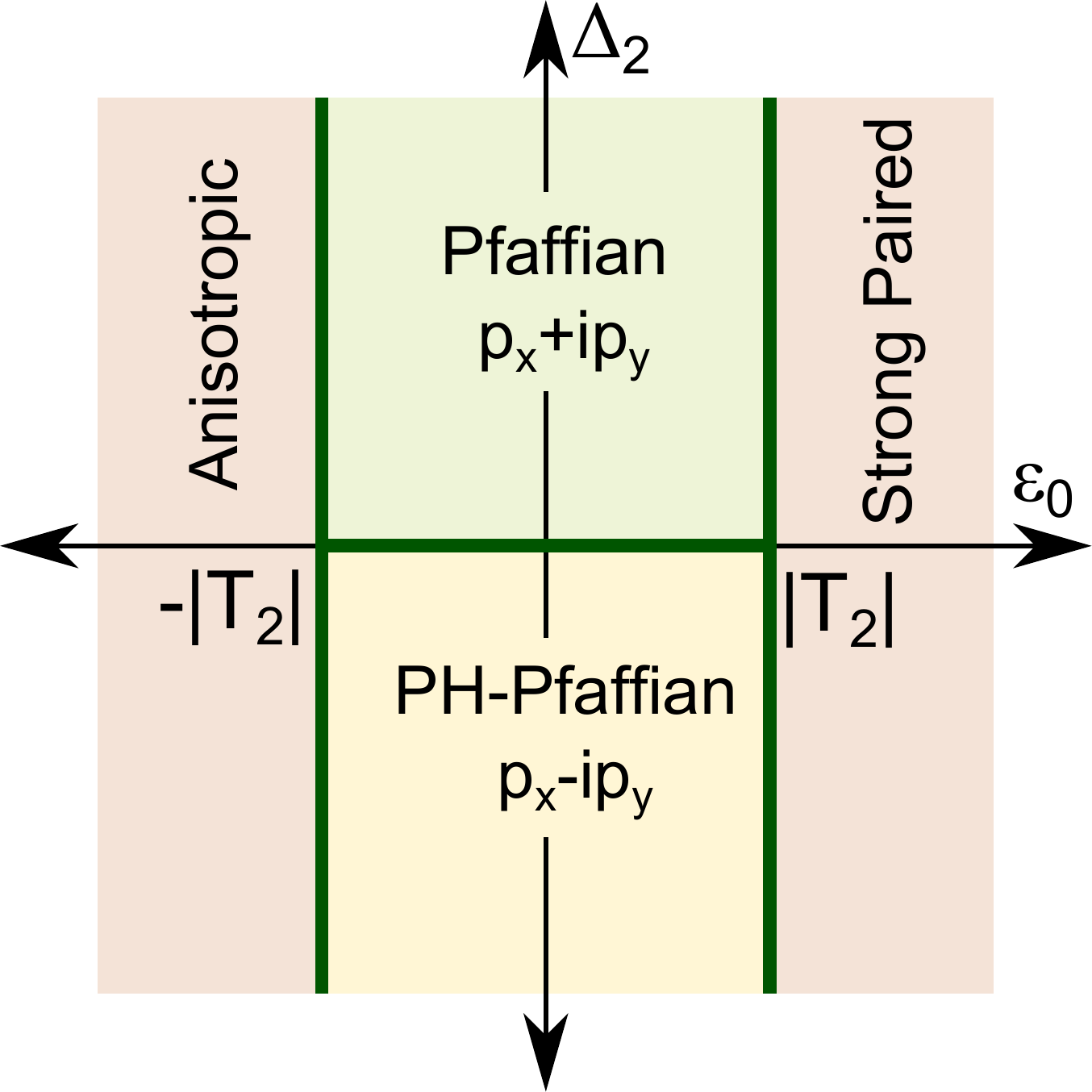}
\caption{Phase diagram for $m$ odd as a function of parameters $\epsilon_0$ and $\Delta_2$ in Eq. \ref{h2dmodd}.  }
\label{phase diagram}
\end{figure}

The  $T_1$ term violates $C_2$ symmetry, and $\Delta_1$ is a $p^y$ pairing term with the same phase as the $p^x$ pairing term on the wires.  To get a gapped phase in the bulk the important pairing term is $\Delta_2$, which gives $\pm i p^y$ pairing.
We consider therefore the simplified model
\begin{equation}
H = \gamma^T [ u k_x \tau^z + (\epsilon_0 + k_x^2/2m + T_2 \cos k_y)\tau^y + \Delta_2 \sin k_y \tau^x]\gamma
\label{h2dmodd}
\end{equation}
As a function of $\epsilon_0$ and $\Delta_2$ this model exhibits several  phases indicated in Fig. \ref{phase diagram} and discussed below.

\subsubsection{Strong Paired Phase}
When $\epsilon_0$ is large and positive each wire is in a strong paired phase, so that the resulting phase is simply a strong paired quantum Hall state of charge $2e$ bosons at filling $1/(4m)$.   This phase describes the region of the phase diagram with $\epsilon_0 > |T_2|$.

As was the case when $m$ was odd, it is a quantum Hall state with a bulk energy gap and a gapless chiral charge mode on the boundary.   There is a single particle energy gap everywhere, including the edge.   This phase is pictorially represented in Fig. \ref{wirediagram}(a) for the case where $t_{1,ab}=0$ and $\epsilon_0>0$, where the Majorana modes couple within each wire to open a gap in the neutral sector.

\subsubsection{Pfaffian and PH-Pfaffian Phases}

For $|\epsilon_0|< |T_2|$ the neutral sector fermions exhibit a 2D topological phase with a non zero Chern number $N = {\rm sgn}  \ \Delta_2$.    The neutral sector has the structure of a weakly paired $p_x\pm ip_y$ superconductor, and there exists a chiral Majorana edge mode, in addition to the charge mode at the edge (which is identical to that in the strong paired phase).

Interestingly, there are two distinct phases depending on the sign of $\Delta_2$, which correspond to   interactions that produce either a $p_x+ ip_y$ phase or a $p_x-ip_y$ phase in the neutral sector.   The $p_x+ip_y$ state has the same topological order as the Moore-Read Pfaffian state, which has copropagating charge and neutral modes at the edge which lead to a chiral central charge $c=1 + 1/2 = 3/2$.
In contrast, for the $p_x-ip_y$ state, the neutral Majorana mode propagates in the opposite direction from the charge mode, leading to a chiral central charge $c= 1- 1/2 = 1/2$.  This state has the same topological order as the variant of the Moore-Read state, dubbed the PH-Pfaffian\cite{son2015}, that has recently been discussed in connection with the particle-hole symmetric half filled Landau level. Though our coupled wire model does not have particle-hole symmetry, the topological order of this state is compatible with particle-hole symmetry.

\begin{figure}
\includegraphics[width=3in]{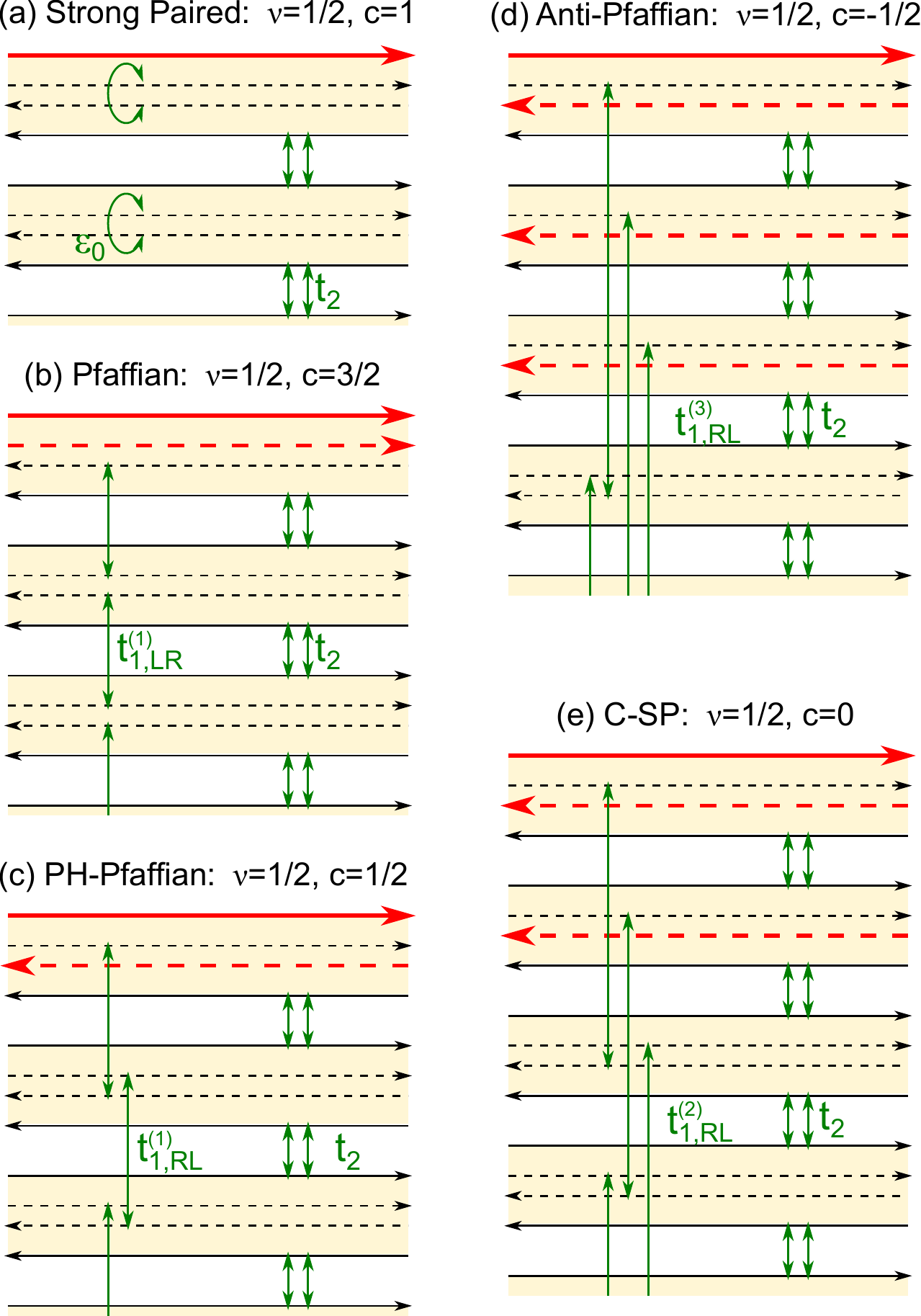}
\caption{Schematic diagrams for quantum Hall states at $\nu=1/2$.   Each shaded region depicts a wire at the pairing transition with right and left moving Majorana modes (dashed lines) as well as right and left moving charge modes (solid lines) describing $\phi_{\rho, i}^{R/L} = \varphi_{\rho,i} \pm m \theta_{\rho,i}$ with $m=2$.   The charge modes are coupled by $t_2$ in (\ref{t2equation}), leaving a single unpaired chiral charge mode on each edge.  The Majorana modes are coupled in different ways described in the text, resulting in topologically distinct states with different numbers of chiral Majorana edge modes.  In (d) C-SP refers to the particle-hole conjugate of the strong paired state.}
\label{wirediagram}
\end{figure}

These states can most easily be pictured in the limit $\epsilon_0=0$ and only $t_{1,RL}$ (or $t_{1,LR}$) is non zero, shown in Fig. \ref{wirediagram}(b,c).  In this case, the chiral Majorana modes on neigboring wires couple and open a gap leaving a single chiral Majorana mode at the edge.

\subsubsection{Anti-Pfaffian Phase and Generalizations}

Though it is outside of the scope of the simple model in Eqs. (\ref{h000i}-\ref{h000f}), the coupled wire model also allows the construction of additional paired quantum Hall states if further neighbor interactions are included.    Here we will keep the pair tunneling in the charge sector to be given by $t_2$ in (\ref{t2equation}), and consider the effect of further neighbor hopping  of single electrons.

A momentum conserving electron tunneling term that connects wire $j$ to wire $j+p$ can be written as
\begin{equation}
V_{1j}^{(p)} = \sum_{a,b} t_{1,ab}^{(p)} \gamma_j^a \gamma_{j+p}^b \cos\left(\sum_{l = 1,p} \Theta_{\rho,j+l-1/2}\right).
\end{equation}
This term involves both tunneling of an electron from wire $j$ to wire $j+p$ as well as backcattering of electrons on the wires in between.
When $\Theta_{\rho,j+1/2}$ is pinned by (\ref{t2equation}), this leads to a Hamiltonian for the Majorana fermions $\gamma_j^a$ that couples $p$'th neighbors.    When expressed in momentum space, as in (\ref{h2by2}), ${\bf h}(k)$ can be characterized by a more general Chern number $N$, such that $|N|\le p$.   This leads to phases with $N$ chiral Majorana modes, where $N>0$ ($N<0$) indicates they are right (left) movers.    Taking into account the charge mode, this leads to a quantum Hall state with chiral central charge $c=1 + N/2$.   The cases $N=0$ (strong paired), $N=1$ (Pfaffian) and $N=-1$ (PH-Pfaffian) were discussed above.

The anti-Pfaffian state\cite{levin2007,lee2007} fits into this more general construction.   This state is the particle-hole conjugate of the Pfaffian state, with $\nu=1/2$ and $c = 1 - 3/2 = -1/2$.    This corresponds to Chern number $N=-3$, which is depicted in Fig. \ref{wirediagram}(e) when $t_{1,RL}^{(3)}$ couples right and left moving Majorana modes on third neighbors, leaving three upstream Majorana modes at the edge.

%In principle, there is a hierarchy of different states with all possible values of $N$.   States with even $N$ will be Abelian states.   For instance, the state with $N=-2$, depicted in Fig. \ref{wirediagram}(d), with $c=0$, is topologically equivalent to the particle-hole conjugate of the strong paired state.    We note that particle-hole conjugate states can also have a simpler construction than that presented above, for instance by considering ``anti-wires" inside a $\nu=1$ quantum Hall state.  However, we will not pursue that direction here.

In principle, there is a hierarchy of different states with all possible values of $N$.   States with even $N$ will be Abelian states.   For instance, the state with $N=-2$, depicted in Fig. \ref{wirediagram}(d), with $c=0$, is topologically equivalent to the particle-hole conjugate of the strong paired state.    When $N$ is even, quasiparticles will bind an even number of Majorana zero modes.   In principle, a kink where $t^{(p=|N|)}_{ab}$ changes sign can bind an unpaired Majorana mode.   However, unless there is an extra symmetry that forbids odd-neighbor coupling, $t^{(p={\rm odd})}_{ab}$, the unpaired Majorana modes will in general be confined.
Finally, we note that particle-hole conjugate states can also have an alternative simpler construction by considering ``anti-wires" inside a $\nu=1$ quantum Hall state, or equivalently ``shifted wires" made from the right moving electron modes of wire $i$ paired with the left mover on wire $i+1$.  However, we will not pursue that direction here.

%We could describe quasiparticles, ground state degeneracy, etc. though that simply will explain what everyone already knows.

\subsubsection{Anisotropic Phase}

When $\epsilon_0$ is large and negative we have an anisotropic quantum Hall state that to our knowledge has not been discussed before.   Both the charged and neutral sectors are gapped in the bulk.   However, the neutral sector has the topological structure similar to that of the decoupled limit $t_{ab}\rightarrow 0$.   The neutral sector of each wire has the structure of a 1D topological superconductor, and for $\epsilon_0<-|T_2|$ these 1D topological superconductors are coupled together to form a ``weak topological superconductor".

This phase does not have unpaired Majorana modes bound to the charge $e/2m$ quasiparticles.  However, at the {\it ends} of the wires in the decoupled limit each wire has a Majorana zero mode at the end.   When the wires are then coupled together, these Majorana modes broaden to form a band.
Provided the lattice of wires has the symmetry under translation by one lattice constant, this band of Majorana modes is necessarily gapless.

If we label the Majorana mode on each wire as $\gamma_j$, then the low energy Hamiltonian will be
\begin{equation}
H = \sum_j i t \gamma_j \gamma_{j+1}
\end{equation}
This has precisely the structure of Kitaev's 1D Majorana chain.   The discrete translation symmetry by one lattice constant guarantees that it is precisely at criticality.   The Majorana modes will have
dispersion
\begin{equation}
E_k = t \sin k
\end{equation}
which exhibits a pair of helical Majorana modes at $k=0$ and $k=\pi$. Furthermore, defects in the wires' array, e.g. dislocations, would bind static non-Abelian defects in the form of isolated Majorana modes.

\section{Discussion and Conclusion\label{discussion}}

This paper has developed a framework for incorporating the physics of pairing into the traditional Luttinger liquid formulation of interacting fermions in one dimension.   We argued that for sufficently strong attractive interactions even a single channel Luttinger liquid composed of spinless fermions can exhibit a strong paired phase that is qualitatively distinct from a weakly paired Luttinger liquid.  It is distinguished by the existence of a single particle energy gap, despite the existence of gapless two particle excitations and a gapless collective charge mode.

These two phases of a Luttinger liquid are in one to one correspondence with the topological and trivial superconducting phases of a one dimensional superconductor.   The fluctuating phase of the one dimensional superconductor describes a Luttinger liquid, and the topological and trivial phases of the superconductor are {\it indistinguishable} from the strong and weak paired phases of the Luttinger liquid.   We have shown that the hallmarks of 1D topological superconductors associated with Majorana end modes - including the zero energy peak in the tunneling density of states at the end and the absence of an even-odd effect in the ground state energy - have a natural correspondence in a single channel Luttinger liquid.

It will be interesting to demonstrate this correspondence experimentally in a Rashba nanowire coupled to a one dimensional superconductor.   In that case, by tuning a magnetic field and gate voltage it should be possible to alternate between trivial and topological superconducting phases on the nanowire as even and odd numbers of channels in the nanowire are populated.  Provided the 1D superconductor is not too stiff (so that $K_\rho$ is not too large), then by tunneling into the middle of the wire it should be possible to demonstrate that there is a gap to single particle excitations in the trivial (strongly paired) phase, while there is no gap to single particle excitations in the topological (weakly paired) phase.

We have also found that paired Luttinger liquids offer new insights into the quantum Hall effect using the coupled wire model.   In addition to allowing the formulation of a new (and much simpler) coupled wire model for the Moore-Read state, it allows us to describe a number of additional phases and critical points at filling $\nu = 1/m$, including the particle-hole symmetric PH-Pfaffian phase, an intrinsically anisotropic phase when $m$ is even, and a strongly paired state when $m$ is odd.

$$ \\ $$
\acknowledgments
We thank Ehud Altman for helpful discussions and for pointing us to Ref. \onlinecite{alberton}.  We also thank Nick Read and Chetan Nayak for helpful discussions. This work was supported in part by  grants from the Microsoft Corporation and  the US-Israel Binational Science Foundation (BIH and AS),  the European
Research Council under the European Unions
Seventh Framework Program (FP7/2007-2013) / ERC
Project MUNATOP, the DFG (CRC/Transregio 183, EI
519/7-1), Minerva foundation (AS) and a Simons Investigator grant from the Simons Foundation (CLK).

\appendix

\section{Mapping to Ising Model}
Here we review the fermionization of the 1+1D transverse field Ising model, which demonstrates the identification of the spin operator above.

On a 1D lattice, the transverse field Ising model has Hamiltonian
\begin{equation}
{\cal H}_I = \sum_j  h \sigma_j^z - J \sigma_j^x \sigma_{j+1}^x
\end{equation}
This can be fermionized by introducing lattice fermion operators
\begin{equation}
c_j^\dagger = \sigma_j^+  \prod_{i<j} \sigma_i^z,
\end{equation}
where $\sigma^\pm = (\sigma^x \pm i \sigma^y)/2$.  This leads to a Kitaev-like lattice fermion model,
\begin{equation}
{\cal H}_I  = \sum_n   h(2 c_j^\dagger c_j-1) - J (c_j^\dagger c_{j+1} + c_j c_{j+1} + h.c.)
\end{equation}
or
\begin{equation}
{\cal H}_I = \sum_k  2(h - J \cos k) c_k^\dagger c_k + 2 J \sin k (c_{-k} c_k + h.c.)
\end{equation}
For $h \sim J$ this has a topological transition.   Near that point we can take the continuum limit, expanding around $k=0$ and going to real space we have
\begin{equation}
{\cal H}_I = \psi^\dagger( \epsilon_0 - \partial_x^2/2m)\psi + iu (\psi \partial_x \psi + h.c.)
\end{equation}
with $\epsilon_0 = 2(h-J)$, $1/2m = J$ and $u = 2J$.   This has precisely the form of $H_f$.  Introducing Majorana operators $\psi = \gamma_1 + i \gamma_2$, we may write this as
\begin{equation}
H_I = \gamma^T \left[-i u \tau^z \partial_x + (\epsilon_0 - \partial_x^2/2m) \tau^y \right]\gamma
\end{equation}

By undoing the Jordan Wigner transformation we can express the spin operators in terms of fermions,
\begin{equation}
\sigma_j^+ = c_j^\dagger e^{i \pi \sum_{i>j} c_m^\dagger c_i}
\end{equation}
In the continuum we thus obtain
\begin{equation}
\sigma^+(x) = \psi^\dagger(x) e^{i \pi \int_x^\infty dx' \psi^\dagger\psi}.
\end{equation}

%\begin{acknowledgments}

%\end{acknowledgments}

\bibliography{refs}

\begin{thebibliography}{29}
\expandafter\ifx\csname natexlab\endcsname\relax\def\natexlab#1{#1}\fi
\expandafter\ifx\csname bibnamefont\endcsname\relax
  \def\bibnamefont#1{#1}\fi
\expandafter\ifx\csname bibfnamefont\endcsname\relax
  \def\bibfnamefont#1{#1}\fi
\expandafter\ifx\csname citenamefont\endcsname\relax
  \def\citenamefont#1{#1}\fi
\expandafter\ifx\csname url\endcsname\relax
  \def\url#1{\texttt{#1}}\fi
\expandafter\ifx\csname urlprefix\endcsname\relax\def\urlprefix{URL }\fi
\providecommand{\bibinfo}[2]{#2}
\providecommand{\eprint}[2][]{\url{#2}}

\bibitem[{\citenamefont{Haldane}(1981{\natexlab{a}})}]{haldane81jpc}
\bibinfo{author}{\bibfnamefont{F.~D.~M.} \bibnamefont{Haldane}},
  \bibinfo{journal}{Journal of Physics C: Solid State Physics}
  \textbf{\bibinfo{volume}{14}}, \bibinfo{pages}{2585}
  (\bibinfo{year}{1981}{\natexlab{a}}),
  \urlprefix\url{http://stacks.iop.org/0022-3719/14/i=19/a=010}.

\bibitem[{\citenamefont{Haldane}(1981{\natexlab{b}})}]{haldane81prl}
\bibinfo{author}{\bibfnamefont{F.~D.~M.} \bibnamefont{Haldane}},
  \bibinfo{journal}{Phys. Rev. Lett.} \textbf{\bibinfo{volume}{47}},
  \bibinfo{pages}{1840} (\bibinfo{year}{1981}{\natexlab{b}}),
  \urlprefix\url{http://link.aps.org/doi/10.1103/PhysRevLett.47.1840}.

\bibitem[{\citenamefont{Kane et~al.}(2002)\citenamefont{Kane, Mukhopadhyay, and
  Lubensky}}]{kml2002}
\bibinfo{author}{\bibfnamefont{C.~L.} \bibnamefont{Kane}},
  \bibinfo{author}{\bibfnamefont{R.}~\bibnamefont{Mukhopadhyay}},
  \bibnamefont{and} \bibinfo{author}{\bibfnamefont{T.~C.}
  \bibnamefont{Lubensky}}, \bibinfo{journal}{Phys. Rev. Lett.}
  \textbf{\bibinfo{volume}{88}}, \bibinfo{pages}{036401}
  (\bibinfo{year}{2002}),
  \urlprefix\url{http://link.aps.org/doi/10.1103/PhysRevLett.88.036401}.

\bibitem[{\citenamefont{Teo and Kane}(2014)}]{teokane2014}
\bibinfo{author}{\bibfnamefont{J.~C.~Y.} \bibnamefont{Teo}} \bibnamefont{and}
  \bibinfo{author}{\bibfnamefont{C.~L.} \bibnamefont{Kane}},
  \bibinfo{journal}{Phys. Rev. B} \textbf{\bibinfo{volume}{89}},
  \bibinfo{pages}{085101} (\bibinfo{year}{2014}),
  \urlprefix\url{http://link.aps.org/doi/10.1103/PhysRevB.89.085101}.

\bibitem[{\citenamefont{Neupert et~al.}(2014)\citenamefont{Neupert, Chamon,
  Mudry, and Thomale}}]{neupert2014}
\bibinfo{author}{\bibfnamefont{T.}~\bibnamefont{Neupert}},
  \bibinfo{author}{\bibfnamefont{C.}~\bibnamefont{Chamon}},
  \bibinfo{author}{\bibfnamefont{C.}~\bibnamefont{Mudry}}, \bibnamefont{and}
  \bibinfo{author}{\bibfnamefont{R.}~\bibnamefont{Thomale}},
  \bibinfo{journal}{Phys. Rev. B} \textbf{\bibinfo{volume}{90}},
  \bibinfo{pages}{205101} (\bibinfo{year}{2014}),
  \urlprefix\url{http://link.aps.org/doi/10.1103/PhysRevB.90.205101}.

\bibitem[{\citenamefont{Sagi and Oreg}(2014)}]{sagi2014}
\bibinfo{author}{\bibfnamefont{E.}~\bibnamefont{Sagi}} \bibnamefont{and}
  \bibinfo{author}{\bibfnamefont{Y.}~\bibnamefont{Oreg}},
  \bibinfo{journal}{Phys. Rev. B} \textbf{\bibinfo{volume}{90}},
  \bibinfo{pages}{201102} (\bibinfo{year}{2014}),
  \urlprefix\url{http://link.aps.org/doi/10.1103/PhysRevB.90.201102}.

\bibitem[{\citenamefont{Santos et~al.}(2015)\citenamefont{Santos, Huang, Gefen,
  and Gutman}}]{santos2015}
\bibinfo{author}{\bibfnamefont{R.~A.} \bibnamefont{Santos}},
  \bibinfo{author}{\bibfnamefont{C.-W.} \bibnamefont{Huang}},
  \bibinfo{author}{\bibfnamefont{Y.}~\bibnamefont{Gefen}}, \bibnamefont{and}
  \bibinfo{author}{\bibfnamefont{D.~B.} \bibnamefont{Gutman}},
  \bibinfo{journal}{Phys. Rev. B} \textbf{\bibinfo{volume}{91}},
  \bibinfo{pages}{205141} (\bibinfo{year}{2015}),
  \urlprefix\url{http://link.aps.org/doi/10.1103/PhysRevB.91.205141}.

\bibitem[{\citenamefont{Meng et~al.}(2015)\citenamefont{Meng, Neupert, Greiter,
  and Thomale}}]{meng2015}
\bibinfo{author}{\bibfnamefont{T.}~\bibnamefont{Meng}},
  \bibinfo{author}{\bibfnamefont{T.}~\bibnamefont{Neupert}},
  \bibinfo{author}{\bibfnamefont{M.}~\bibnamefont{Greiter}}, \bibnamefont{and}
  \bibinfo{author}{\bibfnamefont{R.}~\bibnamefont{Thomale}},
  \bibinfo{journal}{Phys. Rev. B} \textbf{\bibinfo{volume}{91}},
  \bibinfo{pages}{241106} (\bibinfo{year}{2015}),
  \urlprefix\url{http://link.aps.org/doi/10.1103/PhysRevB.91.241106}.

\bibitem[{\citenamefont{Sagi and Oreg}(2015)}]{sagi2015}
\bibinfo{author}{\bibfnamefont{E.}~\bibnamefont{Sagi}} \bibnamefont{and}
  \bibinfo{author}{\bibfnamefont{Y.}~\bibnamefont{Oreg}},
  \bibinfo{journal}{Phys. Rev. B} \textbf{\bibinfo{volume}{92}},
  \bibinfo{pages}{195137} (\bibinfo{year}{2015}),
  \urlprefix\url{http://link.aps.org/doi/10.1103/PhysRevB.92.195137}.

\bibitem[{\citenamefont{Huang et~al.}(2016)\citenamefont{Huang, Chen, Gomes,
  Neupert, Chamon, and Mudry}}]{huang2016}
\bibinfo{author}{\bibfnamefont{P.-H.} \bibnamefont{Huang}},
  \bibinfo{author}{\bibfnamefont{J.-H.} \bibnamefont{Chen}},
  \bibinfo{author}{\bibfnamefont{P.~R.~S.} \bibnamefont{Gomes}},
  \bibinfo{author}{\bibfnamefont{T.}~\bibnamefont{Neupert}},
  \bibinfo{author}{\bibfnamefont{C.}~\bibnamefont{Chamon}}, \bibnamefont{and}
  \bibinfo{author}{\bibfnamefont{C.}~\bibnamefont{Mudry}},
  \bibinfo{journal}{Phys. Rev. B} \textbf{\bibinfo{volume}{93}},
  \bibinfo{pages}{205123} (\bibinfo{year}{2016}),
  \urlprefix\url{http://link.aps.org/doi/10.1103/PhysRevB.93.205123}.

\bibitem[{\citenamefont{Iadecola et~al.}(2016)\citenamefont{Iadecola, Neupert,
  Chamon, and Mudry}}]{iadecola2016}
\bibinfo{author}{\bibfnamefont{T.}~\bibnamefont{Iadecola}},
  \bibinfo{author}{\bibfnamefont{T.}~\bibnamefont{Neupert}},
  \bibinfo{author}{\bibfnamefont{C.}~\bibnamefont{Chamon}}, \bibnamefont{and}
  \bibinfo{author}{\bibfnamefont{C.}~\bibnamefont{Mudry}},
  \bibinfo{journal}{Phys. Rev. B} \textbf{\bibinfo{volume}{93}},
  \bibinfo{pages}{195136} (\bibinfo{year}{2016}),
  \urlprefix\url{http://link.aps.org/doi/10.1103/PhysRevB.93.195136}.

\bibitem[{\citenamefont{Moore and Read}(1991)}]{mooreread1991}
\bibinfo{author}{\bibfnamefont{G.}~\bibnamefont{Moore}} \bibnamefont{and}
  \bibinfo{author}{\bibfnamefont{N.}~\bibnamefont{Read}},
  \bibinfo{journal}{Nuclear Physics B} \textbf{\bibinfo{volume}{360}},
  \bibinfo{pages}{362 } (\bibinfo{year}{1991}), ISSN \bibinfo{issn}{0550-3213},
  \urlprefix\url{http://www.sciencedirect.com/science/article/pii/055032139190407O}.

\bibitem[{\citenamefont{Read and Rezayi}(1999)}]{readrezayi1999}
\bibinfo{author}{\bibfnamefont{N.}~\bibnamefont{Read}} \bibnamefont{and}
  \bibinfo{author}{\bibfnamefont{E.}~\bibnamefont{Rezayi}},
  \bibinfo{journal}{Phys. Rev. B} \textbf{\bibinfo{volume}{59}},
  \bibinfo{pages}{8084} (\bibinfo{year}{1999}),
  \urlprefix\url{http://link.aps.org/doi/10.1103/PhysRevB.59.8084}.

\bibitem[{\citenamefont{Son}(2015)}]{son2015}
\bibinfo{author}{\bibfnamefont{D.~T.} \bibnamefont{Son}},
  \bibinfo{journal}{Phys. Rev. X} \textbf{\bibinfo{volume}{5}},
  \bibinfo{pages}{031027} (\bibinfo{year}{2015}),
  \urlprefix\url{http://link.aps.org/doi/10.1103/PhysRevX.5.031027}.

\bibitem[{\citenamefont{Kitaev}(2001)}]{kitaev2001}
\bibinfo{author}{\bibfnamefont{A.~Y.} \bibnamefont{Kitaev}},
  \bibinfo{journal}{Physics-Uspekhi} \textbf{\bibinfo{volume}{44}},
  \bibinfo{pages}{131} (\bibinfo{year}{2001}),
  \urlprefix\url{http://stacks.iop.org/1063-7869/44/i=10S/a=S29}.

\bibitem[{\citenamefont{Sau et~al.}(2011)\citenamefont{Sau, Halperin,
  Flensberg, and Das~Sarma}}]{sau2011}
\bibinfo{author}{\bibfnamefont{J.~D.} \bibnamefont{Sau}},
  \bibinfo{author}{\bibfnamefont{B.~I.} \bibnamefont{Halperin}},
  \bibinfo{author}{\bibfnamefont{K.}~\bibnamefont{Flensberg}},
  \bibnamefont{and}
  \bibinfo{author}{\bibfnamefont{S.}~\bibnamefont{Das~Sarma}},
  \bibinfo{journal}{Phys. Rev. B} \textbf{\bibinfo{volume}{84}},
  \bibinfo{pages}{144509} (\bibinfo{year}{2011}),
  \urlprefix\url{http://link.aps.org/doi/10.1103/PhysRevB.84.144509}.

\bibitem[{\citenamefont{Fidkowski et~al.}(2011)\citenamefont{Fidkowski,
  Lutchyn, Nayak, and Fisher}}]{fidkowski2011}
\bibinfo{author}{\bibfnamefont{L.}~\bibnamefont{Fidkowski}},
  \bibinfo{author}{\bibfnamefont{R.~M.} \bibnamefont{Lutchyn}},
  \bibinfo{author}{\bibfnamefont{C.}~\bibnamefont{Nayak}}, \bibnamefont{and}
  \bibinfo{author}{\bibfnamefont{M.~P.~A.} \bibnamefont{Fisher}},
  \bibinfo{journal}{Phys. Rev. B} \textbf{\bibinfo{volume}{84}},
  \bibinfo{pages}{195436} (\bibinfo{year}{2011}),
  \urlprefix\url{http://link.aps.org/doi/10.1103/PhysRevB.84.195436}.

\bibitem[{\citenamefont{Kraus et~al.}(2013)\citenamefont{Kraus, Dalmonte,
  Baranov, L\"auchli, and Zoller}}]{KrausZoller}
\bibinfo{author}{\bibfnamefont{C.~V.} \bibnamefont{Kraus}},
  \bibinfo{author}{\bibfnamefont{M.}~\bibnamefont{Dalmonte}},
  \bibinfo{author}{\bibfnamefont{M.~A.} \bibnamefont{Baranov}},
  \bibinfo{author}{\bibfnamefont{A.~M.} \bibnamefont{L\"auchli}},
  \bibnamefont{and} \bibinfo{author}{\bibfnamefont{P.}~\bibnamefont{Zoller}},
  \bibinfo{journal}{Phys. Rev. Lett.} \textbf{\bibinfo{volume}{111}},
  \bibinfo{pages}{173004} (\bibinfo{year}{2013}),
  \urlprefix\url{http://link.aps.org/doi/10.1103/PhysRevLett.111.173004}.

\bibitem[{\citenamefont{{Chen} et~al.}(2017)\citenamefont{{Chen}, {Yan},
  {Ting}, {Chen}, and {Burnell}}}]{ChenBurnell}
\bibinfo{author}{\bibfnamefont{C.}~\bibnamefont{{Chen}}},
  \bibinfo{author}{\bibfnamefont{W.}~\bibnamefont{{Yan}}},
  \bibinfo{author}{\bibfnamefont{C.~S.} \bibnamefont{{Ting}}},
  \bibinfo{author}{\bibfnamefont{Y.}~\bibnamefont{{Chen}}}, \bibnamefont{and}
  \bibinfo{author}{\bibfnamefont{F.~J.} \bibnamefont{{Burnell}}},
  \bibinfo{journal}{ArXiv e-prints}  (\bibinfo{year}{2017}),
  \eprint{1701.01794}.

\bibitem[{\citenamefont{Fu and Kane}(2009)}]{fukane2009}
\bibinfo{author}{\bibfnamefont{L.}~\bibnamefont{Fu}} \bibnamefont{and}
  \bibinfo{author}{\bibfnamefont{C.~L.} \bibnamefont{Kane}},
  \bibinfo{journal}{Phys. Rev. B} \textbf{\bibinfo{volume}{79}},
  \bibinfo{pages}{161408} (\bibinfo{year}{2009}),
  \urlprefix\url{http://link.aps.org/doi/10.1103/PhysRevB.79.161408}.

\bibitem[{\citenamefont{Sitte et~al.}(2009)\citenamefont{Sitte, Rosch, Meyer,
  Matveev, and Garst}}]{sitte}
\bibinfo{author}{\bibfnamefont{M.}~\bibnamefont{Sitte}},
  \bibinfo{author}{\bibfnamefont{A.}~\bibnamefont{Rosch}},
  \bibinfo{author}{\bibfnamefont{J.~S.} \bibnamefont{Meyer}},
  \bibinfo{author}{\bibfnamefont{K.~A.} \bibnamefont{Matveev}},
  \bibnamefont{and} \bibinfo{author}{\bibfnamefont{M.}~\bibnamefont{Garst}},
  \bibinfo{journal}{Phys. Rev. Lett.} \textbf{\bibinfo{volume}{102}},
  \bibinfo{pages}{176404} (\bibinfo{year}{2009}),
  \urlprefix\url{http://link.aps.org/doi/10.1103/PhysRevLett.102.176404}.

\bibitem[{\citenamefont{Alberton et~al.}(2016)\citenamefont{Alberton, Ruhman,
  Berg, and Altman}}]{alberton}
\bibinfo{author}{\bibfnamefont{O.}~\bibnamefont{Alberton}},
  \bibinfo{author}{\bibfnamefont{J.}~\bibnamefont{Ruhman}},
  \bibinfo{author}{\bibfnamefont{E.}~\bibnamefont{Berg}}, \bibnamefont{and}
  \bibinfo{author}{\bibfnamefont{E.}~\bibnamefont{Altman}},
  \bibinfo{journal}{arXiv:1609.02599v2}  (\bibinfo{year}{2016}),
  \urlprefix\url{https://arxiv.org/abs/1609.02599}.

\bibitem[{\citenamefont{Kane and Fisher}(1992)}]{kanefisher1991}
\bibinfo{author}{\bibfnamefont{C.~L.} \bibnamefont{Kane}} \bibnamefont{and}
  \bibinfo{author}{\bibfnamefont{M.~P.~A.} \bibnamefont{Fisher}},
  \bibinfo{journal}{Phys. Rev. B} \textbf{\bibinfo{volume}{46}},
  \bibinfo{pages}{15233} (\bibinfo{year}{1992}),
  \urlprefix\url{http://link.aps.org/doi/10.1103/PhysRevB.46.15233}.

\bibitem[{\citenamefont{Halperin}(1983)}]{halperin1983}
\bibinfo{author}{\bibfnamefont{B.~I.} \bibnamefont{Halperin}},
  \bibinfo{journal}{Helvetica Physica Acta} \textbf{\bibinfo{volume}{56}},
  \bibinfo{pages}{75} (\bibinfo{year}{1983}).

\bibitem[{\citenamefont{Overbosch and Wen}(2016)}]{WenOverbosch}
\bibinfo{author}{\bibfnamefont{B.}~\bibnamefont{Overbosch}} \bibnamefont{and}
  \bibinfo{author}{\bibfnamefont{X.}~\bibnamefont{Wen}},
  \bibinfo{journal}{arXiv:0804.2087}  (\bibinfo{year}{2016}),
  \urlprefix\url{https://arxiv.org/abs/0804.2087}.

\bibitem[{\citenamefont{Albrecht et~al.}(2016)\citenamefont{Albrecht,
  Higginbotham, Madsen, Kuemmeth, Jespersen, Nyg{\aa}rd, Krogstrup, and
  Marcus}}]{Albrecht2016}
\bibinfo{author}{\bibfnamefont{S.~M.} \bibnamefont{Albrecht}},
  \bibinfo{author}{\bibfnamefont{A.~P.} \bibnamefont{Higginbotham}},
  \bibinfo{author}{\bibfnamefont{M.}~\bibnamefont{Madsen}},
  \bibinfo{author}{\bibfnamefont{F.}~\bibnamefont{Kuemmeth}},
  \bibinfo{author}{\bibfnamefont{T.~S.} \bibnamefont{Jespersen}},
  \bibinfo{author}{\bibfnamefont{J.}~\bibnamefont{Nyg{\aa}rd}},
  \bibinfo{author}{\bibfnamefont{P.}~\bibnamefont{Krogstrup}},
  \bibnamefont{and} \bibinfo{author}{\bibfnamefont{C.~M.}
  \bibnamefont{Marcus}}, \bibinfo{journal}{Nature}
  \textbf{\bibinfo{volume}{531}}, \bibinfo{pages}{206} (\bibinfo{year}{2016}),
  ISSN \bibinfo{issn}{0028-0836}, \bibinfo{note}{letter},
  \urlprefix\url{http://dx.doi.org/10.1038/nature17162}.

\bibitem[{\citenamefont{Sondhi and Yang}(2001)}]{sondhi2001}
\bibinfo{author}{\bibfnamefont{S.~L.} \bibnamefont{Sondhi}} \bibnamefont{and}
  \bibinfo{author}{\bibfnamefont{K.}~\bibnamefont{Yang}},
  \bibinfo{journal}{Phys. Rev. B} \textbf{\bibinfo{volume}{63}},
  \bibinfo{pages}{054430} (\bibinfo{year}{2001}),
  \urlprefix\url{http://link.aps.org/doi/10.1103/PhysRevB.63.054430}.

\bibitem[{\citenamefont{Levin et~al.}(2007)\citenamefont{Levin, Halperin, and
  Rosenow}}]{levin2007}
\bibinfo{author}{\bibfnamefont{M.}~\bibnamefont{Levin}},
  \bibinfo{author}{\bibfnamefont{B.~I.} \bibnamefont{Halperin}},
  \bibnamefont{and} \bibinfo{author}{\bibfnamefont{B.}~\bibnamefont{Rosenow}},
  \bibinfo{journal}{Phys. Rev. Lett.} \textbf{\bibinfo{volume}{99}},
  \bibinfo{pages}{236806} (\bibinfo{year}{2007}),
  \urlprefix\url{http://link.aps.org/doi/10.1103/PhysRevLett.99.236806}.

\bibitem[{\citenamefont{Lee et~al.}(2007)\citenamefont{Lee, Ryu, Nayak, and
  Fisher}}]{lee2007}
\bibinfo{author}{\bibfnamefont{S.-S.} \bibnamefont{Lee}},
  \bibinfo{author}{\bibfnamefont{S.}~\bibnamefont{Ryu}},
  \bibinfo{author}{\bibfnamefont{C.}~\bibnamefont{Nayak}}, \bibnamefont{and}
  \bibinfo{author}{\bibfnamefont{M.~P.~A.} \bibnamefont{Fisher}},
  \bibinfo{journal}{Phys. Rev. Lett.} \textbf{\bibinfo{volume}{99}},
  \bibinfo{pages}{236807} (\bibinfo{year}{2007}),
  \urlprefix\url{http://link.aps.org/doi/10.1103/PhysRevLett.99.236807}.

\end{thebibliography}

\end{document}